\documentclass[12pt]{article}
\usepackage{amsfonts}
\usepackage{amsmath}
\usepackage{amssymb}
\usepackage{amsthm}
\usepackage[all]{xy} 
\usepackage{verbatim} 
\usepackage[bookmarks = true, colorlinks=true, linkcolor = black, citecolor = black, menucolor = black, urlcolor = black]{hyperref}
\usepackage{tikz}
\usetikzlibrary{arrows}
\usetikzlibrary{arrows.new}

\newtheorem{Thm}{Theorem}[section]

\newtheorem{Cor}[Thm]{Corollary}

\theoremstyle{definition}
\newtheorem{Defn}[Thm]{Definition}
\newtheorem{Rem}[Thm]{Remark}

\newtheorem{Note}[Thm]{Note}

\newtheoremstyle{algorithm}{3ex}{\topsep}{}{}{\bfseries}{.}{\newline}{}
\theoremstyle{algorithm}
\newtheorem{Alg}{Algorithm}

\newcommand{\Zset}{\mathbb{Z}}         

\newcommand{\Nset}{\mathbb{N}}         

\DeclareMathOperator{\im}{Im} 
\DeclareMathOperator{\kr}{Ker} 
\DeclareMathOperator{\rank}{rank}
\DeclareMathOperator{\card}{card}
\DeclareMathOperator{\pers}{pers}

\DeclareMathOperator{\inc}{inc}
\DeclareMathOperator{\dgm}{Dgm}

\newcommand{\algitem}{\begin{list}{$\bullet$}{\setlength{\topsep}{0pt}}
  \setlength{\itemsep}{0pt}}

\newcommand{\inp}{{\itshape Input: }}
\newcommand{\outp}{{\itshape Output: }}

\newcommand{\rrdc}{\mbox{\,\(\Rightarrow\hspace{-9pt}\Rightarrow\)\,}}
\newcommand{\lrdc}{\mbox{\,\(\Leftarrow\hspace{-9pt}\Leftarrow\)\,}}
\newcommand{\lrrdc}{\mbox{\,\(\Leftarrow\hspace{-9pt}\Leftarrow\hspace{-5pt}\Rightarrow\hspace{-9pt}\Rightarrow\)\,}} 

\newcommand{\coordsb}[2]{\foreach \i in {0,...,#1} \foreach \j in {0,...,#2}
                         {\coordinate (\i\j) at (\i,\j) ;
                          \node (n\i\j) at (\i\j) {\scriptsize\(\bullet\)} ;}}

\title{Defining and computing persistent $\Zset$-homology in the general case}
\author{Ana Romero, J\'onathan Heras,  Julio Rubio, and Francis Sergeraert}
\date{}

\begin{document}
\maketitle

\begin{abstract}
By general case we mean methods able to process simplicial sets and chain complexes not of finite type.
A filtration of the object to be studied is the heart of both subjects persistent homology and spectral sequences. In this paper we present
the complete relation between them, both
from theoretical and computational points of
view. One of the main contributions of this paper is the observation
that a slight modification of our previous programs
computing spectral sequences is enough to compute also
persistent homology. By inheritance from our spectral
sequence programs, we obtain for free persistent homology
programs applicable to spaces not of finite type (provided they
are spaces with effective homology) and with $\Zset$-coefficients
(significantly generalizing the usual presentation of persistent homology over
a field). As an illustration, we compute some persistent homology
groups (and the corresponding integer barcodes) in the case
of a Postnikov tower.
\end{abstract}


\section{Introduction}

Persistent homology is an algebraic method for measuring topological
features of shapes and functions, with many recent applications such as point cloud data \cite{Ghr08}, sensor networks \cite{SG07}, optical character recognition \cite{Ked10} and protein classification \cite{Zom05}. The main underlying idea of all these applications is that relevant features will be \emph{long-lived} in the sense that they persist over a certain parameter range, on contrast with the ``noise'' which will be \emph{short-lived}.

There exist several algorithms designed for computing the persistent homology groups $H^{i,j}_n$ of simplicial complexes, see for instance \cite{EH08} and \cite{ZC05}. However, these techniques can only be applied when the considered space is finite, and up to our knowledge there is no method for computing persistence in the infinite case (that is to say, computing persistence of infinite dimensional spaces, but with homology groups of finite type).

Spectral sequences are a tool in Algebraic Topology providing information on the homology of a complex by means of successive approximations, and they are known to be related with persistent homology (see \cite{EH10} and \cite{ZC05}). In a previous work \cite{RRS06}, a set of programs was developed for computing spectral sequences of filtered complexes, allowing the computation of the different components of the spectral sequence even if the filtered complex is not of finite type, if the \emph{effective homology} \cite{RS02} of the complex is known.

The goal of this work consists in using our programs for spectral sequences for computing persistent homology of filtered complexes. Since our spectral sequence programs are able to work in the general case, the results about persistent homology will also be valid for simplicial sets not of finite type. Moreover, we generalize the usual computations of persistent homology over a field by defining and computing persistent homology with $\Zset$-coefficients, where extension problems can be found. A slight modification of our algorithms for spectral sequences has been sufficient to allow us to determine the persistent homology groups $H^{i,j}_n$.

Our calculations have made it possible to detect an error in \cite{EH10}: the so called ``Spectral sequence theorem'' (\cite[p. 171]{EH10}) claims that:

\begin{quote}\emph{The total rank of the groups of dimension $p+q$ in the level $r \geq 1$ of the associated spectral sequence equals the number of points
in the (p + q)-th persistence diagram whose persistence is $r$ or larger, that~is,
$$
\sum_{p=1}^m \rank E^r_{p,q} = \card \{a \in \dgm_{p+q}(f)| \pers(a) \geq r\}
$$}\end{quote}

However, our programs have showed that the above result is false: in the left side of the formula there can be more elements than in the right side and the relation is in fact an inequality.

The paper is organized as follows. After this brief introduction to our problem, we present the main contributions of this paper in Section~\ref{sec:main_contributions}. Section~\ref{sec:persistent-homology} presents a generalization of the usual definition of persistent homology over a field by defining persistent homology with $\Zset$-coefficients. In Section~\ref{sec:pers-hom-and-ss-relation}, we explain the relation between persistent homology and spectral sequences, showing that the result in \cite{EH10} is false and providing the correct connection between both concepts. In Section~\ref{sec:spct-seq-infinite} we present some programs for computing spectral sequences of filtered complexes, which can be used in the general case of spaces which are not of finite type. These programs can be applied for the computation of persistent $\Zset$-homology as explained in Section~\ref{sec:computing-prst}. We then include some examples of interesting applications in Section~\ref{sec:examples_and_applications}. The paper ends with a section of conclusions.

\section{Main contributions of the paper}
\label{sec:main_contributions}

This work includes different new contributions. In this section, we enumerate them in order to make the paper more understandable.

\begin{enumerate}

\item The paper presents a new module for the Kenzo system \cite{Kenzo} allowing the user to compute persistent homology groups, enhancing in this way the functionality of the program.

\item The use of this new module has made it possible to detect an error in a theorem published in \cite{EH10}: the so called ``Spectral sequence theorem'' \cite[p. 171]{EH10}, relating the notions of persistent homology and spectral sequences, includes a formula which is not correct.

\item A new correct formula explaining the relation between persistent homology and spectral sequences is stated. Here, we provide an expression for determining the persistent homology groups of a filtered complex in terms of the associated spectral sequence; see \cite{BP13} for a different formula where spectral sequence groups are defined in terms of persistent homology.

\item Our new Kenzo programs computing persistent homology groups consider the general case of $\Zset$-coefficients. This allows for example to differentiate the persistence of some concrete filtrations of a Klein bottle and a torus (which have indistinguishable persistent homology groups over $\Zset_2$).

\item Moreover, our programs can also deal with spaces of infinite type, and are able to produce for instance persistent homology groups of complicated stages of Postnikov towers.

\item The persistent homology groups determined by our programs are provided with generators. This opens the possibility of a \emph{qualitative} study of persistent homology, going beyond the traditional \emph{quantitative} analysis (based, for instance, on barcodes). With our approach, we can trace the born and death moments of particular cycles, and their contribution to the persistent homology groups.

\end{enumerate}

All these results have been obtained using the \emph{effective homology} method \cite{RS02}, a technique for computing homology groups of complicated (infinite) spaces. The effective homology technique is based
on the following idea: given some topological spaces $X_1, \ldots,
X_n$, a topological constructor $\Phi$ produces a new topological
space~$X$. If effective homology versions of the spaces $X_1,
\ldots, X_n$ are known, then we should be able to build an effective homology
version of the space $X$, and such a version would allow us to
compute the homology groups of $X$.

A typical example of this kind of situation is the loop space
constructor. Given a $1$-reduced simplicial set $X$ with effective
homology, it is possible to determine the effective homology of
the loop space $\Omega(X)$, which in particular allows us to
compute the homology groups $H_\ast(\Omega(X))$. Moreover, if $X$
is \mbox{$m$-reduced}, this process may be iterated $m$ times, producing
an effective homology version of $\Omega^k(X)$, for $k \leq m$.
Effective homology versions are also
known for classifying spaces or total spaces of fibrations, see
\cite{RS06} for more information.

\section{Persistent homology in the general case}
\label{sec:persistent-homology}

\subsection{Preliminaries}
\label{subsec:pers-hom-preliminaries}
Let us begin by introducing some basic definitions and results about persistent homology. For details, see \cite{EH10}.

\begin{Defn}
\label{defn:sc-filtration}
Let $K$ be a simplicial complex. A (finite) \emph{filtration} of $K$ is a nested sequence of subcomplexes \(K^i \subseteq K\) such that
\[
\emptyset= K^{0} \subseteq K^1 \subseteq K^{2} \subseteq \cdots \subseteq K^{m}=K
\]
\end{Defn}

For every $i \leq j$ we have an inclusion map on the canonically associated chain complexes $\inc^{i,j}: C(K^i) \hookrightarrow C(K^j)$ and therefore we can consider the induced homomorphisms $f^{i,j}_n : H_n(K^i) \rightarrow H_n(K^j)$, for each dimension $n$. The filtration produces then for each dimension $n$ a sequence of homology groups connected by homomorphisms:

$$
0=H_n(K^0) \rightarrow H_n(K^1) \rightarrow \cdots \rightarrow H_n(K^m) =H_n(K)
$$

As we go from $K^{i-1}$ to $K^i$, we gain new homology classes (corresponding to cycles which are in $C(K^i)$ but not in $C(K^{i-1})$) and we lose some classes when they become trivial or merge with other classes. 

\begin{Defn}
\label{defn:persistent-homology}
The $n$-th \emph{persistent homology groups} of $K$, denoted by $H^{i,j}_n(K)\equiv H^{i,j}_n$, are the images of the homomorphisms $f^{i,j}_n$:
$$ H^{i,j}_n = \im f^{i,j}_n, \mbox{ for } 0 \leq i \leq j \leq m$$

The group $H^{i,j}_n$ consists of the $n$-th homology classes of $K^i$ that are still alive at $K^j$. A class $\gamma \in H_n(K^i)$ is said to \emph{be born} at $K^i$ if $\gamma \notin H^{i-1,i}_n$. It is said to \emph{die} entering $K^j$ if it merges with an older class as we go from $K^{j-1}$ to $K^j$, that is, $f^{i,j-1}_n(\gamma) \notin H^{i-1,j-1}_n$ but $f^{i,j}_n(\gamma) \in H^{i-1,j}_n$. If $\gamma$ is born at $K^i$ and dies entering $K^j$, the difference $j-i$ is called the \emph{persistence index} of $\gamma$, denoted $\pers(\gamma)$. If $\gamma$ is born at $K^i$ but never dies then $\pers(\gamma)=\infty$.
\end{Defn}

If the homology is computed with field coefficients, each group $H^{i,j}_n$ is a vector space which is determined up to isomorphism by its dimension. We denote $\beta^{i,j}_n:=\rank H^{i,j}_n$. This allows us to represent all persistent homology groups in a visual way by means of a \emph{barcode} diagram, defined as follows.

\begin{Defn}
A \emph{$P$-interval} is a half-open interval $[i , j)$, which is also represented by its endpoints
$(i , j) \in \Zset \times (\Zset \cup \infty)$. A class that is born at $K^i$ and never dies is represented as $(i,\infty)$. A class that is born at $K^i$ and dies entering $K^j$ is represented as $(i , j)$. Finite multisets of $P$-intervals are plotted as disjoint unions of intervals, called \emph{barcodes}.
\end{Defn}

\begin{figure}[h]
\centering

\begin{tikzpicture}

\draw (0,0) node[rectangle,draw ]{
\begin{tikzpicture}

\draw (0,0) node[rectangle,draw]{
\begin{tikzpicture}[scale=0.5]
\coordinate (0) at (0,0);
\coordinate (1) at (2,0);
\coordinate (2) at (1,2);
\draw[fill=black] (0) circle (1pt and 1pt);
\draw[fill=white,color=white] (1) circle (1pt and 1pt);
\draw[fill=white,color=white] (2) circle (1pt and 1pt);
\end{tikzpicture}};

\draw (1.75,0) node[rectangle,draw ]{
\begin{tikzpicture}[scale=0.5]
\coordinate (0) at (0,0);
\coordinate (1) at (2,0);
\coordinate (2) at (1,2);
\draw[fill=black] (0) circle (1pt and 1pt);
\draw[fill=black] (1) circle (1pt and 1pt);
\draw[fill=white,color=white] (2) circle (1pt and 1pt);

\end{tikzpicture}};

\draw (3.5,0) node[rectangle,draw ]{
\begin{tikzpicture}[scale=0.5]
\coordinate (0) at (0,0);
\coordinate (1) at (2,0);
\coordinate (2) at (1,2);
\draw[fill=black] (0) circle (1pt and 1pt);
\draw[fill=black] (1) circle (1pt and 1pt);
\draw[fill=black] (2) circle (1pt and 1pt);
\end{tikzpicture}};

\draw (5.25,0) node[rectangle,draw ]{
\begin{tikzpicture}[scale=0.5]
\coordinate (0) at (0,0);
\coordinate (1) at (2,0);
\coordinate (2) at (1,2);
\draw (0) -- (1);
\draw[fill=black] (0) circle (1pt and 1pt);
\draw[fill=black] (1) circle (1pt and 1pt);
\draw[fill=black] (2) circle (1pt and 1pt);
\end{tikzpicture}};

\draw (7,0) node[rectangle,draw ]{
\begin{tikzpicture}[scale=0.5]
\coordinate (0) at (0,0);
\coordinate (1) at (2,0);
\coordinate (2) at (1,2);
\draw (0) -- (1);
\draw (1) -- (2);
\draw[fill=black] (0) circle (1pt and 1pt);
\draw[fill=black] (1) circle (1pt and 1pt);
\draw[fill=black] (2) circle (1pt and 1pt);
\end{tikzpicture}};

\draw (8.75,0) node[rectangle,draw ]{
\begin{tikzpicture}[scale=0.5]
\coordinate (0) at (0,0);
\coordinate (1) at (2,0);
\coordinate (2) at (1,2);
\draw (0) -- (1);
\draw (1) -- (2);
\draw (0) -- (2);
\draw[fill=black] (0) circle (1pt and 1pt);
\draw[fill=black] (1) circle (1pt and 1pt);
\draw[fill=black] (2) circle (1pt and 1pt);
\end{tikzpicture}};

\draw (10.5,0) node[rectangle,draw ]{
\begin{tikzpicture}[scale=0.5]
\coordinate (0) at (0,0);
\coordinate (1) at (2,0);
\coordinate (2) at (1,2);
\draw (0) -- (1);
\draw (1) -- (2);
\draw (0) -- (2);
\draw[fill=black] (0) circle (1pt and 1pt);
\draw[fill=black] (1) circle (1pt and 1pt);
\draw[fill=black] (2) circle (1pt and 1pt);
\draw[fill=gray!80!white] (0)--(1)--(2)--cycle;
\end{tikzpicture}};

\end{tikzpicture}
};
\end{tikzpicture}
\caption{filtration of a triangle in seven steps.}\label{fig:triangle-filtration}
\end{figure}
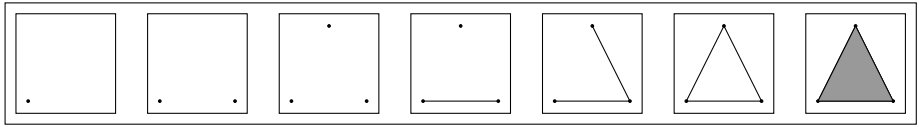

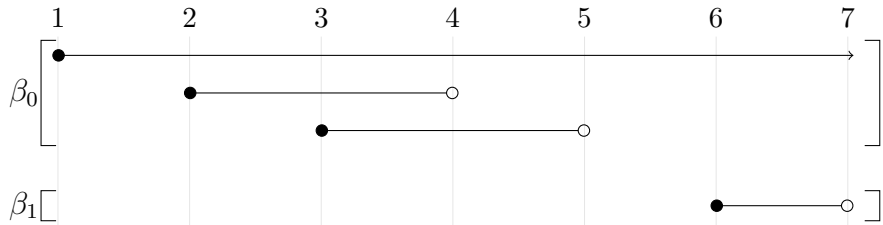
\begin{figure}[h]
\centering
\begin{tikzpicture}

\draw[step=1,gray!20!white,very thin,fill=white] (-5.35,-1.25) -- (-5.35,-3.8);
\draw (-5.35,-1) node {{\small $1$}};

\draw[step=1,gray!20!white,very thin,fill=white] (-3.6,-1.25) -- (-3.6,-3.8);
\draw (-3.6,-1) node {{\small $2$}};

\draw[step=1,gray!20!white,very thin,fill=white] (-1.85,-1.25) -- (-1.85,-3.8);
\draw (-1.85,-1) node {{\small $3$}};

\draw[step=1,gray!20!white,very thin,fill=white] (-0.1,-1.25) -- (-0.1,-3.8);
\draw (-0.1,-1) node {{\small $4$}};

\draw[step=1,gray!20!white,very thin,fill=white] (1.65,-1.25) -- (1.65,-3.8);
\draw (1.65,-1) node {{\small $5$}};

\draw[step=1,gray!20!white,very thin,fill=white] (3.4,-1.25) -- (3.4,-3.8);
\draw (3.4,-1) node {{\small $6$}};

\draw[step=1,gray!20!white,very thin,fill=white] (5.15,-1.25) -- (5.15,-3.8);
\draw (5.15,-1) node {{\small $7$}};

\draw (-5.8,-2) node {$\beta_0$};
\draw (0,-2) node{
\begin{tikzpicture}
\draw (-.7,1.2) -- (-.9,1.2) -- (-.9,-.2) -- (-.7,-.2);
\draw [*->] (-.75,1) -- (9.9,1);
\draw [*-o] (1,.5) -- (4.65,.5);
\draw [*-o] (2.75,0) -- (6.4,0);
\draw (10.05,1.2) -- (10.25,1.2) -- (10.25,-.2) -- (10.05,-.2);

\end{tikzpicture}
};

\draw (-5.8,-3.5) node {$\beta_1$};
\draw (0,-3.5) node{
\begin{tikzpicture}
\draw (-.7,0.2) -- (-.9,0.2) -- (-.9,-.2) -- (-.7,-.2);
\draw [*-o] (8,0) -- (9.9,0);
\draw (10.05,0.2) -- (10.25,.2) -- (10.25,-.2) -- (10.05,-.2);

\end{tikzpicture}
};
\end{tikzpicture}
\caption{barcode diagram of the filtered triangle.}\label{fig:triangle-barcodes}
\end{figure}

For instance, let us consider a triangle with seven filtration steps as described in Figure \ref{fig:triangle-filtration} and let us study the homology groups of the different stages of the filtration. In dimension $0$, there are three non-null persistent homology classes: one is born at $K^1$ and never dies; the second one is born at $K^2$ and dies entering $K^4$; and the third one is born at $K^3$ and dies entering $K^5$. In dimension $1$, we have only one persistent class which is born at $K^6$ and dies entering $K^7$. In dimension $2$, there are not non-null persistent homology groups. All these classes can be represented by means of barcodes as in Figure \ref{fig:triangle-barcodes}.

\subsection{Integral persistence}
\label{subsec:integer-persistence}

The main references on persistent homology \cite{EH08}, \cite{EH10} or \cite{ZC05} consider the initial case of coefficients in a field, so that the representation by means of a barcode diagram as explained in the previous subsection totally describes the persistent homology groups $H^{i,j}_n$. We focus now on the universal integer case (the universal coefficient theorem \cite{Hat02} explains how the $\Zset$ case contains as a by-product any other particular case), where one can face extension problems. This section includes new results and ideas providing a generalization of the classical persistent homology over a field to the case of integer coefficients.


Let us begin by observing that the groups $H^{\ast,j}_n$ (Definition \ref{defn:persistent-homology}) provide a filtration of $H^j_n\equiv H_n(K^j)$:
\begin{equation}
\label{eq:persistent-filtration}
0=H ^{0,j}_n \subseteq H^{1,j}_n \subseteq H^{2,j}_n \subseteq \cdots \subseteq H^{j,j}_n=H^j_n
\end{equation}

We can also consider a \emph{double filtration} of $H^j_n$ obtained by introducing the \emph{new} groups $H^{i,j,k}_n$, for $i \leq j \leq k$, defined as
$$
H^{i,j,k}_n=H^{i,j}_n \cap (f_n^{j,k})^{-1}(H^{i-1,k}_n) \subseteq H^j_n
$$

For each fixed $i$ and $j$, the different groups $H^{i,j,\ast}_n$ define a filtration between $H^{i-1,j}_n$ and $H^{i,j}_n$:
\begin{equation}
\label{eq:persistent-filtration-2}
H^{i-1,j}_n= H_n^{i,j,j} \subseteq H^{i,j,j+1}_n \subseteq  H^{i,j,j+2}_n \subseteq \cdots \subseteq H^{i,j,m}_n \subseteq  H^{i,j}_n
\end{equation}

Each group $H^{i,j,k}_n$ contains all classes which are in $H^{i-1,j}_n$ and also the classes of $H^j_n$ which are born at $K^i$ and die at or before $K^k$.

An interesting property of this filtration is given by the following fact: when we apply the map $f^{j,j+1}_n$ to the elements of the different subgroups $H^{i,j,k}_n$ (that is, we consider the elements of $H^{i,j,k}_n$, which are homology classes of $H^j_n$, as classes in  $H^{j+1}_n$), we obtain the relation
$$f^{j,j+1}_n(H^{i,j,k}_n) = H^{i,j+1,k}_n\subseteq H^{i,j+1}_n\subseteq H^{j+1}_n$$

\begin{figure}[h]
\centering
\fbox{
 \begin{tikzpicture} [xscale = 2.5, yscale = -2, inner sep = -4pt, anchor = -20, font = \scriptsize]
 \coordsb{4}{5}
  \node at (0,-0.12) {0} ;
  \node at (1,-0.12) {0} ;
  \node at (2,-0.12) {0} ;
  \node at (3,-0.12) {0} ;
  \node at (4,-0.12) {0} ;
  \node at (0,0.88) {\(H_n^{1,1,2}\)} ;
  \node at (0,1.88) {\(H_n^{1,1,3}\)} ;
  \node at (0,2.88) {\(H_n^{1,1,4}\)} ;
  \node at (0,3.88) {\(H_n^{1,1,5}\)} ;
  \node at (0,4.88) {\(H_n^{1,1}\)} ;
  \draw [-angle 45 new, arrow head=3mm] (01) -- (10);
  \draw [-angle 45 new, arrow head=3mm]  (02) -- (11);
   \draw [-angle 45 new, arrow head=3mm] (11) --(20);
   \draw [-angle 45 new, arrow head=3mm] (03) -- (12);
    \draw [-angle 45 new, arrow head=3mm] (12)-- (21);
     \draw [-angle 45 new, arrow head=3mm](21) -- (30);
    \draw [-angle 45 new, arrow head=3mm] (04) -- (13);
     \draw [-angle 45 new, arrow head=3mm] (13) -- (22);
      \draw [-angle 45 new, arrow head=3mm] (22)-- (31);
       \draw [-angle 45 new, arrow head=3mm] (31) -- (40) ;
   \draw [-angle 45 new, arrow head=3mm]  (05) -- (14);
    \draw [-angle 45 new, arrow head=3mm] (14) -- (23);
     \draw [-angle 45 new, arrow head=3mm] (23)-- (32);
      \draw [-angle 45 new, arrow head=3mm] (32) -- (41);
 \draw [-angle 45 new, arrow head=3mm]        (15) -- (24);
  \draw [-angle 45 new, arrow head=3mm] (24) -- (33);
   \draw [-angle 45 new, arrow head=3mm] (33) --  (42) ;
  \draw [-angle 45 new, arrow head=3mm] (25) -- (34);
   \draw [-angle 45 new, arrow head=3mm] (34) --  (43);
   \draw [-angle 45 new, arrow head=3mm] (35) -- (44) ;
 \node at (1,4.25) {\scriptsize\(\bullet\)} ;
 \node at (1,4.5) {\scriptsize\(\bullet\)} ;
 \node at (1,4.75) {\scriptsize\(\bullet\)} ;
 \node at (2,3.333) {\scriptsize\(\bullet\)} ;
 \node at (2,3.667) {\scriptsize\(\bullet\)} ;
 \node at (2,4.333) {\scriptsize\(\bullet\)} ;
 \node at (2,4.667) {\scriptsize\(\bullet\)} ;
 \node at (3,2.5) {\scriptsize\(\bullet\)} ;
 \node at (3,3.5) {\scriptsize\(\bullet\)} ;
 \node at (3,4.5) {\scriptsize\(\bullet\)} ;
 \draw[-angle 45 new, arrow head=3mm] (1,4.25) to [out = -50, in = 115] (2,3) ;
 \draw[-angle 45 new, arrow head=3mm] (1,4.5) -- (2,3.333);
   \draw[-angle 45 new, arrow head=3mm] (2,3.333) to [out = -50, in = 105] (3,2) ;
 \draw[-angle 45 new, arrow head=3mm] (1,4.75) -- (2,3.667) ;
  \draw[-angle 45 new, arrow head=3mm]  (2, 3.667) -- (3,2.5) ;
   \draw[-angle 45 new, arrow head=3mm]  (3,2.4) to [out = -50, in = 105] (4,1) ;
 \draw[-angle 45 new, arrow head=3mm] (2,4.333) to [out = -50, in = 105] (3,3) ;
 \draw[-angle 45 new, arrow head=3mm] (2,4.667) -- (3,3.5) ;
  \draw[-angle 45 new, arrow head=3mm]  (3,3.5) to [out = -50, in = 105] (4,2) ;
 \draw[-angle 45 new, arrow head=3mm] (3,4.5) to [out = -50, in = 105] (4,3) ;
 \node at (1,0.88) {\(H_n^{1,2,3}\)} ;
 \node at (1,1.88) {\(H_n^{1,2,4}\)} ;
 \node at (2,0.88) {\(H_n^{1,3,4}\)} ;
 \node at (1,2.88) {\(H_n^{1,2,5}\)} ;
 \node at (2,1.88) {\(H_n^{1,3,5}\)} ;
 \node at (3,0.88) {\(H_n^{1,4,5}\)} ;
 \node at (1,3.88) {\(H_n^{1,2}\)} ;
 \node at (1,4.13) {\(H_n^{2,2,3}\)} ;
 \node at (1,4.38) {\(H_n^{2,2,4}\)} ;
 \node at (1,4.63) {\(H_n^{2,2,5}\)} ;
 \node at (1,4.88) {\(H_n^{2,2}\)} ;
 \node at (2,2.88) {\(H_n^{1,3}\)} ;
 \node at (2,3.213) {\(H_n^{2,3,4}\)} ;
 \node at (2,3.547) {\(H_n^{2,3,5}\)} ;
 \node at (2,3.88) {\(H_n^{2,3}\)} ;
 \node at (2,4.213) {\(H_n^{3,3,4}\)} ;
 \node at (2,4.547) {\(H_n^{3,3,5}\)} ;
 \node at (2,4.88) {\(H_n^{3,3}\)} ;
 \node at (3,1.88) {\(H_n^{1,4}\)} ;
 \node at (3,2.38) {\(H_n^{2,4,5}\)} ;
 \node at (3,2.88) {\(H_n^{2,4}\)} ;
 \node at (3,3.38) {\(H_n^{3,4,5}\)} ;
 \node at (3,3.88) {\(H_n^{3,4}\)} ;
 \node at (3,4.38) {\(H_n^{4,4,5}\)} ;
 \node at (3,4.88) {\(H_n^{4,4}\)} ;
 \node at (4,0.88) {\(H_n^{1,5}\)} ;
 \node at (4,1.88) {\(H_n^{2,5}\)} ;
 \node at (4,2.88) {\(H_n^{3,5}\)} ;
 \node at (4,3.88) {\(H_n^{4,5}\)} ;
 \node at (4,4.88) {\(H_n^{5,5}\)} ;
 \node at (0.45,0.45) {\(f_n^{1,2}\)} ;
 \node at (1.45,0.45) {\(f_n^{2,3}\)} ;
  \node at (2.45,0.45) {\(f_n^{3,4}\)} ;
   \node at (3.45,0.45) {\(f_n^{4,5}\)} ;
 \end{tikzpicture}%
 }
\caption{relations in the double filtration of persistent homology groups.}\label{fig:double-filtration-relations}
\end{figure}
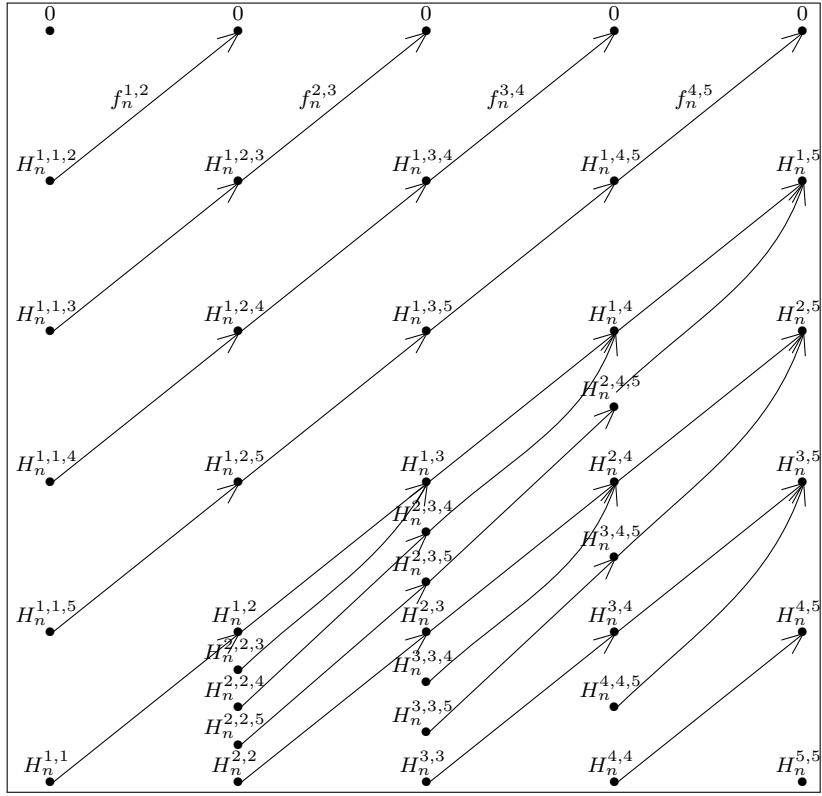

In particular, for $k=j+1$, one has $f^{j,j+1}_n(H^{i,j,j+1}_n) = H^{i,j+1,j+1}_n= H^{i-1,j+1}_n\subseteq H^{j+1}_n$.
We can represent these interesting relations by means of Figure \ref{fig:double-filtration-relations}, where each arrow from $H^{i,j,k}_n$ to $H^{i,j+1,k}_n$ denotes the fact that  $f^{j,j+1}_n(H^{i,j,k}_n) = H^{i,j+1,k}_n$.

This double filtration defines two associated graded modules. The first one, associated to the filtration (\ref{eq:persistent-filtration}), is given by $\widehat{H}^{i,j}_n:=H^{i,j}_n/H^{i-1,j}_n$. The second one, coming from the second filtration (\ref{eq:persistent-filtration-2}), is defined by $\widehat{H}^{i,j,k}_n:=H^{i,j,k}_n/H^{i,j,k-1}_n$.

As before, the morphisms $f^{j,j+1}_n$ induce maps $\widehat{H}^{i,j}_n \rightarrow \widehat{H}^{i,j+1}_n$ and $\widehat{H}^{i,j,k}_n \rightarrow \widehat{H}^{i,j+1,k}_n$. By the very definition, $f^{j,j+1}_n: \widehat{H}^{i,j,k}_n \rightarrow \widehat{H}^{i,j+1,k}_n$ is an isomorphism except for $k=j+1$ where the map is null.


Elementary  linear algebra  proves that the  following quotient  groups are
canonically  isomorphic. We denote  by $BD_n^{i,k}(K)\equiv BD_n^{i,k}$  their common
isomorphism class:

\begin{equation*}
\begin{aligned}
BD_n^{i,k}  = & \frac{H_n^{i,i,k}}{H_n^{i,i,k-1}}  = \frac{H_n^{i,i+1,k}}{H_n^{i,i+1,k-1}} = \cdots \\
              & \cdots =\frac{H_n^{i,k-2,k}}{H_n^{i,k-2,k-1}}= \frac{H_n^{i,k-1,k}}{H_n^{i-1,k-1}}
\end{aligned}
\end{equation*}

 \noindent The notation  $BD_n^{i,k}$ should be read as \emph{the  group of homological
classes \underline{b}orn  at time i and \underline{d}ying  at time k},
in  fact  a group  of  equivalence  classes  modulo inferior  homology
groups.

Each group $BD_n^{i,k}$ admits a canonical divisor presentation:
\[
BD_n^{i,k} \cong \Zset_{d_n^{i,k,1}} \oplus \cdots \oplus \Zset_{d_n^{i,k,p_{n,i,k}}}
\]
every $\Zset$-index $\in {0} \cup [2, \ldots]$ dividing the next one.

When working over a field, the groups $BD^{i,k}_n$ of \mbox{$n$-dimensional} classes that are born at $K^i$ and die entering $K^k$ are uniquely determined (up to isomorphism) by their rank, $\mu^{i,k}_n$, which is given by the formula \cite[p. 152]{EH10}:
$$
\mu^{i,k}_n = (\beta^{i,k-1}_n - \beta^{i,k}_n) - (\beta_n^{i-1,k-1} - \beta_n^{i-1,k})
$$

In this way, the groups $BD^{i,k}_n$ can be determined (up to isomorphism) if the groups $H^{i,j}_n$ are known. Conversely, in the field situation the information about the ranks of $BD^{i,k}_n$ is sufficient to know the \emph{total} groups $H^{i,j}_n$ and also the groups $H^{i,j,k}_n$ defined in our double filtration.

But in the integer coefficient case the situation is not so favorable. Now the groups $H^{i,j}_n$ or $H^{i,j,k}_n$  are not sufficient to determine $BD^{i,k}_n$, because there could be several possibilities for the corresponding quotients\footnote{For example, the quotient of $\Zset_2 \oplus \Zset_4$ by $\Zset_2$ could be $\Zset_2 \oplus \Zset_2$ or $\Zset_4$.}. Similarly, from the groups $BD_n^{i,k}$ it is not always possible to determine the persistent homology groups $H^{i,j}_n$ and $H^{i,j,k}_n$ because of extension problems.

Let us suppose for example that in some dimension $n$ there is a class $\gamma_1$ which is born at $K^1$ and dies entering $K^6$, and another class $\gamma_2$ which is born at $K^3$ and dies at $K^5$, and the groups $BD_n^{1,6}$ and $BD_n^{3,5}$, generated by $\gamma_1$ and $\gamma_2$ respectively, are equal to $\Zset_2$. Both elements $\gamma_1$ and $\gamma_2$ will be in the group $H_n^{3,4}$ of classes which are in $H_n(K^3)$ and still live at $K^4$, but in principle one cannot know if the group $H_n^{3,4}$ is equal to $\Zset_4$ or $\Zset_2 \oplus \Zset_2$, the two possible extensions of $\Zset_2$ by $\Zset_2$. 
If we want to completely understand the persistent structure of a simplicial complex $K$, it is necessary then to determine \emph{separately} the groups $BD^{i,k}_n$, $H^{i,j,k}_n$ and $H^{i,j}_n$. We will solve this problem in Section \ref{sec:computing-prst}.

On the other hand, let us observe that if $K$ is a simplicial set not of finite type, even if the homology groups $H_n(K)$ are of finite type it may happen that some groups $H_n(K^j)$ and $H^{i,j}_n \subseteq H_n(K^j)$ are not of finite type. 

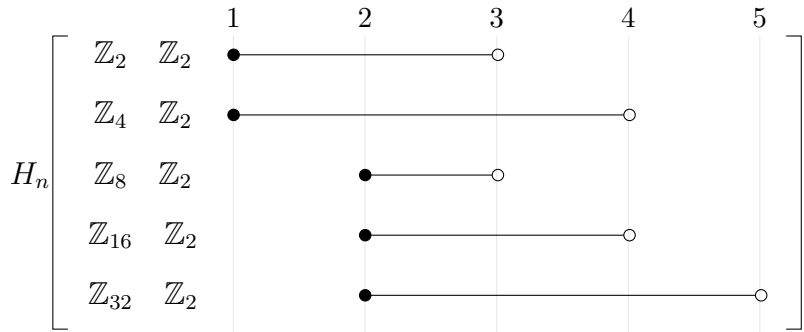
\begin{figure}[h]
\centering
\begin{tikzpicture}


\draw[step=1,gray!20!white,very thin,fill=white] (-3.6,-1.25) -- (-3.6,-5.2);
\draw (-3.6,-1) node {{\small $1$}};

\draw[step=1,gray!20!white,very thin,fill=white] (-1.85,-1.25) -- (-1.85,-5.2);
\draw (-1.85,-1) node {{\small $2$}};

\draw[step=1,gray!20!white,very thin,fill=white] (-0.1,-1.25) -- (-0.1,-5.2);
\draw (-0.1,-1) node {{\small $3$}};

\draw[step=1,gray!20!white,very thin,fill=white] (1.65,-1.25) -- (1.65,-5.2);
\draw (1.65,-1) node {{\small $4$}};

\draw[step=1,gray!20!white,very thin,fill=white] (3.4,-1.25) -- (3.4,-5.2);
\draw (3.4,-1) node {{\small $5$}};

\draw (-6.3,-3.1) node {$H_n$};

\draw (-5.8,-1.25) -- (-6,-1.25) -- (-6,-5.15) -- (-5.8,-5.15);
\draw (3.75,-1.25) -- (3.95,-1.25) -- (3.95,-5.15) -- (3.75,-5.15);

\draw [*-o] (-3.68,-1.5) -- (0,-1.5);
\draw (-4.8,-1.5) node{$\mathbb{Z}_2 \quad \mathbb{Z}_2$};

\draw [*-o] (-3.68,-2.3) -- (1.75,-2.3);
\draw (-4.8,-2.3) node{$\mathbb{Z}_4 \quad \mathbb{Z}_2$};

\draw [*-o] (-1.93,-3.1) -- (0,-3.1);
\draw (-4.8,-3.1) node{$\mathbb{Z}_8 \quad \mathbb{Z}_2$};

\draw [*-o] (-1.93,-3.9) -- (1.75,-3.9);
\draw (-4.8,-3.9) node{$\mathbb{Z}_{16} \quad \mathbb{Z}_2$};

\draw [*-o] (-1.93,-4.7) -- (3.5,-4.7);
\draw (-4.8,-4.7) node{$\mathbb{Z}_{32} \quad \mathbb{Z}_2$};



\end{tikzpicture}
\caption{example of integer barcode diagram.}\label{fig:integer-groupcode-2}
\end{figure}

The  integer  framework,  thanks   to  the  given  definitions  of $H^{i,j,k}_n$ and $BD^{i,k}_n$, leads us to the obvious adaptation of
    the classical barcode diagram. It is based on the different stages of the filtration of $H^j_n$ given by the groups $H^{i,j}_n$ and $H^{i,j,k}_n$. In this description, a bar is included for each stage $H^{i,j,k}_n$ of the filtration which is different from the previous one $H^{i,j,k-1}_n$; it is represented as the interval $[i,k)$, indicating the group $H^{i,j,k}_n$ and also the corresponding quotient by the previous group $H^{i,j,k-1}_n$ (since, as explained before, in the integer case both groups do not always determine the quotient). These diagrams describe completely the persistent structure of a simplicial set; and from them, one can deduce not only the groups $H^j_n$, $H^{i,j}_n$ and $H^{i,j,k}_n$, but also the groups $BD_n^{i,k}$ which represent the homology classes which are born at time $i$ and die at time $k$. The barcode diagram in Figure \ref{fig:integer-groupcode-2} describes a situation with $H_n(K^2)=\Zset_{32}$ filtered~by:
$$
\xymatrix @R=3.0mm @C=-0.5mm{
 H^{0,2}_n \ar@{=}[d] & \subset & H^{1,2,3}_n \ar@{=}[d] &  \subset & H^{1,2,4}_n=H^{1,2}_n \ar@{=}[d] & \subset & H^{2,2,3}_n \ar@{=}[d] & \subset & H^{2,2,4}_n \ar@{=}[d] & \subset & H^{2,2,5}_n=H^{2,2}_n=H^2_n \ar@{=}[d] \\
0 & \subset & \Zset_2 & \subset  & \Zset_4 & \subset & \Zset_8 & \subset & \Zset_{16} & \subset & \Zset_{32}
}
$$

\begin{figure}[h]
\centering
\begin{tikzpicture}

\draw[step=1,gray!20!white,very thin,fill=white] (-5.35,-1.25) -- (-5.35,-2.8);
\draw (-5.35,-1) node {{\small $1$}};

\draw[step=1,gray!20!white,very thin,fill=white] (-3.6,-1.25) -- (-3.6,-2.8);
\draw (-3.6,-1) node {{\small $2$}};

\draw[step=1,gray!20!white,very thin,fill=white] (-1.85,-1.25) -- (-1.85,-2.8);
\draw (-1.85,-1) node {{\small $3$}};

\draw[step=1,gray!20!white,very thin,fill=white] (-0.1,-1.25) -- (-0.1,-2.8);
\draw (-0.1,-1) node {{\small $4$}};

\draw[step=1,gray!20!white,very thin,fill=white] (1.65,-1.25) -- (1.65,-2.8);
\draw (1.65,-1) node {{\small $5$}};

\draw[step=1,gray!20!white,very thin,fill=white] (3.4,-1.25) -- (3.4,-2.8);
\draw (3.4,-1) node {{\small $6$}};

\draw (-6.3,-2) node {$H_n$};

\draw (-5.8,-1.25) -- (-6,-1.25) -- (-6,-2.75) -- (-5.8,-2.75);
\draw (3.75,-1.25) -- (3.95,-1.25) -- (3.95,-2.75) -- (3.75,-2.75);

\draw [*-] (-5.43,-1.5) -- (-2.2,-1.5);
\draw (-5.3,-1.5) node[anchor=north]{$\mathbb{Z}_2$};

\draw [*-o] (-1.93,-2.3) -- (1.74,-2.3);
\draw (-1.8,-2.3) node[anchor=north]{$\mathbb{Z}_2$};

\draw [-o] (1.74,-2.3) -- (3.49,-2.3);
\draw (2,-2.3) node[anchor=north]{$\mathbb{Z}_2$};

\draw[rounded corners=5pt] (-2.2,-1.5) -- (-2,-1.5) --  (-2,-1.7);
\draw (-2.003,-1.7) -- node[anchor=west]{$\mathbb{Z}_4$} (-2.003,-2);
\draw[rounded corners=5pt] (-2,-2) -- (-2,-2.3) --  (-1.8,-2.3);

\end{tikzpicture}
\caption{alternative description for an integer barcode diagram.}\label{fig:integer-barcode}
\end{figure}
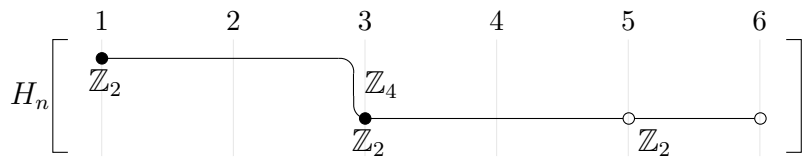

In some simple situations, an alternative description could also be useful. Here an interval $[i,k)$ represents a class which is born at time $i$ and dies entering $K^k$ (as in the barcode diagram when working over a field), but now we will indicate for each interval in the diagram the group $BD_n^{i,k}$ that generates. If some classes in two groups $BD_n^{i,k}$ produce a non-trivial extension, we will join the corresponding intervals indicating also the new group. On the other hand, if only some elements of a component of a group die at $K^k$ we will indicate the quotient. The situation explained before where $BD_n^{1,6}$ and $BD_n^{3,5}$, generated by $\gamma_1$ and $\gamma_2$ respectively, are equal to $\Zset_2$, and $H_n^{3,4}=\Zset_4$, will be represented by means of Figure \ref{fig:integer-barcode}. This alternative description is a generalization of the classical barcode diagram over a field.

The new definitions of groups $B^{i,k}_n$ and $H^{i,j,k}_n$ and the barcode diagram in the integer case generalizes the classical theory of persistent homology over a field. In Section \ref{sec:computing-prst}, we will give algorithms for computing these elements.

\section{Relation between persistent homology and spectral sequences}
\label{sec:pers-hom-and-ss-relation}

Once we have presented the main ideas about persistent homology over a field and we have given the corresponding definitions for the integer case, let us introduce now the notion of spectral sequence associated with a filtered chain complex. For details, see \cite{Mac63}.

\begin{Defn}
A (finite) \emph{filtration} of a chain complex $C$ is a family of sub-chain complexes $C^i \subseteq C$  such that
\[
0= C^0  \subseteq C^1 \subseteq C^2 \subseteq \cdots \subseteq C^m=C
\]
\end{Defn}

\begin{Defn}
Let $R$ be a ring, a \emph{spectral sequence} $E=(E^r,d^r)_{r \geq 1}$ is a sequence of bigraded \mbox{$R$-modules} $E^r=\{E^r_{p,q}\}_{p,q\in \Zset}$, each provided with a differential $d^r=\{d^r_{p,q}: E^r_{p,q}\rightarrow E^r_{p-r, q+r-1}\}_{p,q \in \Zset}$ of bidegree $(-r,r-1)$ (satisfying $d_{p-r,q+r-1} \circ d_{p,q}=0$) and with isomorphisms $H(E^r, d^r)\cong E^{r+1}$ for every $r\geq 1$. Since each $E^{r+1}_{p,q}$ is a subquotient of $E^{r}_{p,q}$, one can define the \emph{final groups} $E^\infty_{p,q}$ of the spectral sequence as the groups which remain after the computation of all successive homologies.
\end{Defn}

\begin{Thm} \cite[p.327]{Mac63}
\label{thm:fltr-spsq}
Let $C$ be a chain complex with a filtration. There exists a spectral sequence $E \equiv E(C) \equiv (E^r, d^r)_{r \geq 1}$, defined by
\[
E^r_{p,q}=\frac{Z^r_{p,q}+ C^{p-1}_{p+q}}{d_{p+q+1}(Z^{r-1}_{p+r-1,q-r+2})+ C^{p-1}_{p+q}}
\]
where $Z^r_{p,q}$ is the submodule $Z^r_{p,q}=\{ a\in C^p_{p+q} | \ d_{p+q}(a)\in C^{p-r}_{p+q-1}\}\subseteq C^p_{p+q}$, and $d^r_{p,q}:E^r_{p,q}\rightarrow E^r_{p-r,q+r-1}$ is the morphism induced on these subquotients by the differential map $d_{p+q}:C_{p+q}\rightarrow C_{p+q-1}$.
This spectral sequence \emph{converges} to the homology groups of $C$, that is,  there are natural isomorphisms
$$
E^\infty_{p,q} \cong \frac{H^p_{p+q}(C)}{H^{p-1}_{p+q}(C)}
$$
where $H^p_\ast(C)$ is the filtration on the homology groups $H_\ast(C)$ induced by the filtration of $C$.
\end{Thm}

Given a simplicial complex $K$, a filtration on $K$ (Definition \ref{defn:sc-filtration}) induces a filtration on the associated chain complex $C(K)$, and therefore it makes sense to consider the associated spectral sequence (defined in Theorem \ref{thm:fltr-spsq}), which converges to the homology groups $H_n(C(K))\equiv H_n(K)$.  Since both persistent homology and spectral sequences provide information about the homology groups of the complex $K$ by means of the homology of some subcomplexes produced by the filtration, it seems natural to study if both objects are related.

As we will see in the next section, in a previous work \cite{RRS06} we have developed some algorithms and programs for computing spectral sequences of filtered chain complexes with $\Zset$-coefficients. The programs (accessible at \cite{Rom06}) are written in Common Lisp, enhancing the Kenzo system \cite{Kenzo}, and allowing the computation of spectral sequences associated with filtered chain complexes, even in some cases of chain complexes of non-finite nature with \emph{effective homology}. The groups $E^r_{p,q}$ of the spectral sequence are determined by means of a function called \texttt{spsq-group}. Although our programs were designed to deal with complicated (not of finite type) spaces, they can also be applied to finite cases. For example, for the spectral sequence associated with the filtration of a triangle as described in Figure \ref{fig:triangle-filtration}, we obtain that the groups (of dimension $p+q=0$) $E^1_{1,-1}$, $E^1_{2,-2}$ and $E^1_{3,-3}$ are equal to $\Zset$:

{\footnotesize \begin{verbatim}
> (spsq-group triangle 1 1 -1)
Spectral sequence E^1_{1,-1}
Component Z
> (spsq-group triangle 1 2 -2)
Spectral sequence E^1_{2,-2}
Component Z
> (spsq-group triangle 1 3 -3)
Spectral sequence E^1_{3,-3}
Component Z
\end{verbatim}}

\normalsize
In dimension $p+q=1$, one has also three non-null groups $E^1_{4,-3}$, $E^1_{5,-4}$ and $E^1_{6,-5}$:

{\footnotesize \begin{verbatim}
> (spsq-group triangle 1 4 -3)
Spectral sequence E^1_{4,-3}
Component Z
> (spsq-group triangle 1 5 -4)
Spectral sequence E^1_{5,-4}
Component Z
> (spsq-group triangle 1 6 -5)
Spectral sequence E^1_{6,-5}
Component Z
\end{verbatim}}

\normalsize

 There are some works in the literature which include some comments on the relation between spectral sequences and persistent homology (see for instance \cite{EH08} and \cite{ZC05}). Moreover, the book ``Computational Topology: An Introduction'' by
Herbert Edelsbrunner and John Harer \cite{EH10} includes an explicit formula relating both concepts. Given a filtered simplicial complex $K$, the so called ``Spectral sequence theorem'' (\cite[p. 171]{EH10}) claims that:

\begin{quote}\emph{
The total rank of the groups of dimension $p+q$ in the level $r \geq 1$ of the associated spectral sequence equals the number of points
in the $(p + q)$-th persistence diagram whose persistence is $r$ or larger, that~is,
\begin{equation*}
\begin{aligned}
\sum_{p=1}^m \rank E^r_{p,q} & = \card \{a \in \dgm_{p+q}(f)| \pers(a) \geq r\}
\end{aligned}
\end{equation*}
where in the left side $q$ decreases as $p$ increases so that the dimension $p+q$ remains constant.}
\end{quote}

According to the definition of \emph{persistence diagram} \cite{EH10}, the right term of the formula, $\card \{a \in \dgm_{p+q}(f)| \pers(a) \geq r\}$, corresponds to the sum of the ranks of all groups $BD^{i,k}_n$ with $k-i \geq r$.

However, considering the example of the triangle of Figures \ref{fig:triangle-filtration} and \ref{fig:triangle-barcodes}, we can observe that there is a discrepancy in the previous formula: for dimension $p+q=1$, there are three groups $E^1_{4,-3}$, $E^1_{5,-4}$ and $E^1_{6,-5}$ equal to~$\Zset$, but the only non-null group $BD^{i,k}_1$ for $k-i \geq 1$ is $BD^{6,7}_1=\Zset$.

Thus, we have detected that the formula in \cite{EH10} is erroneous. We can observe in the definition of the spectral sequence of a filtered complex (Theorem \ref{thm:fltr-spsq}) that each class in $E^r_{p,q}$ is generated by an ``almost'' cycle of dimension $p+q$: a class in $E^r_{p,q}$ is given by a chain whose boundary in~$K^p-K^{p-r}$ is empty but it may have non-empty boundary in $K^{p-r}$. Moreover, the fact of making the quotient by $d_{p+q+1}(Z^{r-1}_{p+r-1,q-r+2})$ and $C^{p-1}_{p+q}$ implies that a (non-null) class of $E^r_{p,q}$ is given by an element $x\in Z^r_{p,q}$ which is not in $K^{p-1}$ nor in the differential of $K^{p+r-1}$. In particular, if $x$ is a cycle (that is, $d (x)=0$), then~$x$ defines a class of $H_{p+q}(K)$ which is born at $K^p$ and is still alive at $K^{p+r-1}$, and then the persistence index of $x$ is at least $r$.

Then, it is worth remarking that in the spectral sequence side (the left side) of the formula in \cite{EH10} there can be more elements than in the persistence (right) side, corresponding to classes in $E^r_{p,q}$ which are ``almost'' cycles but whose boundary is not null (and therefore they do not correspond to classes in the homology groups $H_{p+q}(K^i)$). To illustrate the error in \cite{EH10}, it suffices to consider as a counterexample a simplicial complex $K$ generated by the interval $ab$, with the filtration given by $K^1=\{a, b\}$
and $K^2=K$; in dimension $1$ we have $E^1_{2,-1}=\Zset$ but there are no classes of persistence at least $1$ since the unique element of dimension $1$ is not a cycle.

The formula relating spectral sequences and persistent homology in \cite{EH10} should be in fact an inequality:
\begin{equation*}
\begin{aligned}
\sum_{p=1}^m \rank E^r_{p,q} & \geq  \card \{a \in \dgm_{p+q}(f)| \pers(a) \geq r\}
\end{aligned}
\end{equation*}

One can observe then that the elements of $E^r_{p,q}$ which are also in the corresponding persistent homology groups $H^{i,i+r-1}_{p+q}$ are the elements which are real cycles (that is, their boundary is null). More concretely, the intervals of length equals to $r$ correspond to the images of the differential
maps in the level $E^r$ of the spectral sequence. That is, a class of dimension
$p+q$ which is born at $K^p$ and dies entering $K^{p+r}$ corresponds to a component in
the image of the differential map $d^r_{p+r,q-r+1}:E^r_{p+r,q-r+1} \rightarrow E^r_{p,q}$.
A class of dimension $p+q$ which is born at $K^p$ but never dies corresponds to a
component in the final group in the spectral sequence $E^{\infty}_{p,q}$.

We define:
$$
A^r_{p,q} := \im(d^r_{p+r,q-r+1} : E^r_{p+r,q-r+1} \rightarrow E^r_{p,q})  \subseteq E^r_{p,q}
$$
and we obtain the following theorem expressing the relation between persistent homology and the spectral sequence associated with a filtered simplicial complex.
\begin{Thm}
\label{thm:relation_correct}
Let $ K^1 \subseteq K^{2} \subseteq \cdots \subseteq K^{m}=K$ be a filtered simplicial complex. For each dimension $n$ and for each $0 \leq i < k \leq m$, one has a canonical isomorphism
$$
BD_n^{i,k}\cong A^{k-i}_{i,n-i}= \im(d^{k-i}_{k,n-k+1})
$$
\end{Thm}

\begin{Cor}
\label{cor:relation_correct}
Let $ K^1 \subseteq K^{2} \subseteq \cdots \subseteq K^{m}=K$ be a filtered simplicial complex. The total rank of the images of the differential maps in the level $r\geq 1$ of the spectral sequence equals the sum of the ranks of groups $BD^{i,k}_n$ with $k-i=r$:
$$
\sum_{p=1}^m \rank A^r_{p,q} = \sum_{p=1}^{m-r} \rank BD^{p,r+p}_{p+q}
$$
\end{Cor}

Theorem \ref{thm:relation_correct} and Corollary \ref{cor:relation_correct} explain the relation between persistent homology and spectral sequences and give us an algorithm for computing the  persistent homology groups $BD^{i,k}_n$ of a filtered simplicial complex from the associated spectral sequence. Therefore, if the spectral sequence groups $E^r_{p,q}$ are known with the corresponding generators (and are finitely generated groups so that one can compute the images of the differential maps) then it is also possible to determine the groups $BD^{i,k}_n$, which will be given by a list of generators and the corresponding Betti number and torsion coefficients.

Let us emphasize that if we work with coefficients over a field $F$, knowing the groups $BD_n^{i,k}$ provides also (up to isomorphism) the groups $H^{i,j}_n$. However,
 in the integer case from the groups $BD_n^{i,k}$ it is not always possible to determine the persistent homology groups $H^{i,j}_n$ and the filtration given by $H^{i,j,k}_n$, since extension problems could happen. In Section~\ref{sec:computing-prst}, we will provide algorithms for computing not only the groups $BD_n^{i,k}$ obtained by means of the relation with the spectral sequence, but also the \emph{total} persistent groups $H^{i,j}_n$ and the filtration by $H^{i,j,k}_n$, solving in this way the possible extension problems.

In \cite{BP13}, a different formula relating spectral sequences and persistent homology is stated, expressing the sum of the ranks of the groups $E^r_{p,q}$ in terms of $\beta^{i,j}_n$ and $\mu^{i,k}_n$.

\section{Spectral sequences in the general case}
\label{sec:spct-seq-infinite}

As we have seen in Theorem \ref{thm:fltr-spsq}, the groups $E^r_{p,q}$ of the spectral sequence of a filtered chain complex $C$ are defined by means of a formal expression obtained as a subquotient of some subgroups of $C^p$. However, we can observe that the subgroups which appear in this formal expression ($Z^r_{p,q}$, $C^{p-1}_{p+q}$ and $d_{p+q+1}(Z^{r-1}_{p+r-1,q-r+2})$) can only be determined in an elementary way when the chain complex $C$ is finitely generated in each dimension. Therefore, this formal definition of the different components of the spectral sequence does not provide in general an algorithm for computing the desired groups $E^r_{p,q}$ and  the differential maps~$d^r_{p,q}$.

The \emph{effective homology} method \cite{RS02} provides algorithms for computing homology groups of complicated (infinite) spaces; and, as we will see in this section, it can also be used to determine the different components of the spectral sequence of a filtered chain complex, even in the general case of spaces not of finite type. We begin by introducing the main definitions of this technique which we will use later for computing spectral sequences; for details see \cite{RS02} or \cite{RS06}. All chain complexes considered in this section are chain complexes of free $\Zset$-modules.

\begin{Defn}
\label{def:red}
A \emph{reduction} $\rho\equiv(D \rrdc C)$ between two
chain complexes is a triple $(f,g,h)$ where: (a) The components
$f$ and $g$ are chain complex morphisms $f: D \rightarrow C$ and $g: C \rightarrow D$; (b)
The component $h$ is a homotopy operator $h:D\rightarrow D$ (a graded group homomorphism of degree +1); (c) The following relations are satisfied:
  (1) $f  g = \mbox{id}_C$; (2)
  $g f + d_D h + h  d_D
        = \mbox{id}_D$;
  (3)~\ {$f  h = 0$;} (4) $h   g = 0$; (5) $h   h = 0$.

\end{Defn}

\begin{Rem}
These relations express that $D$ is the direct sum of $C$ and a contractible (acyclic) complex. This decomposition is simply $D=\kr  f \oplus \im  g$, with $\im  g\cong C$ and $H_n(\kr  f )=0$ for all $n$. In particular, this implies that the graded homology groups \(H_\ast(D)\) and \(H_\ast(C)\) are canonically isomorphic.
\end{Rem}

\begin{Defn}
A \emph{(strong chain) equivalence} between the complexes $C$ and~$E$ (denoted by $C \lrrdc E$) is a triple $(D,\rho,\rho ')$ where $D$ is a chain complex, $\rho$ and $\rho'$ are reductions from $D$ over $C$ and  $E$ respectively: $C \stackrel{\rho}{\lrdc} D \stackrel{\rho'}{\rrdc} E.$
\end{Defn}

\begin{Note}
An effective chain complex is essentially a free chain complex $C$ where each group $C_n$ is finitely generated, and there is an algorithm that returns a $\Zset$-base in each degree $n$ (for details, see \cite{RS02}). The homology groups of an effective chain complex $C$ can be determined by means of diagonalization algorithms on matrices \cite{KMM04}.
\end{Note}

\begin{Defn}
An \emph{object with effective homology} is a triple $(X,EC,\varepsilon)$ where $EC$ is an effective chain complex and $\varepsilon$ is an equivalence between a free chain complex canonically associated to $X$ and $EC$, $C(X) \stackrel{\epsilon}{\lrrdc} EC$.
\end{Defn}

\begin{Note}
It is important to understand that in general the \(EC\) component of an object with effective homology is \emph{not} made of the homology groups of~\(X\); this component \(EC\) is a free \(\Zset\)-chain complex of finite type, in general with a non-null  differential, allowing to \emph{compute} the homology groups of \(X\); the justification is the equivalence \(\varepsilon\).

The notion of object with effective homology makes it possible to compute homology groups of complicated spaces by means of homology groups of effective complexes (which can be obtained using diagonalization algorithms on matrices). This method is based on the following idea: given some topological spaces $X_1, \ldots, X_n$, a topological constructor $\Phi$ produces a new topological space $X$. If effective homology versions of the spaces $X_1, \ldots, X_n$ are known, then an effective homology version of the space $X$ can also be built, and this version allows us to compute the homology groups of $X$.
\end{Note}

The effective homology method is implemented in a system called Kenzo~\cite{Kenzo}, a Lisp 16,000 lines program devoted to Symbolic Computation in Algebraic Topology, implemented by the fourth author of this paper and some coworkers. Kenzo works with rich and complex algebraic structures (chain complexes, differential graded algebras, simplicial sets, simplicial groups, morphisms between these objects, reductions, etc.) and has obtained some results (for example homology groups of iterated loop spaces of a loop space modified by a cell attachment, components of complex Postnikov towers, homotopy groups of suspended classifying spaces, etc.) which had never been determined before \cite{RS06}. Moreover, Kenzo has made it possible to detect an error in a theorem published in \cite{MW10}, where some theoretical reasonings are used to deduce that the fourth homotopy group of the suspended classifying space of the fourth alternating group $A_4$, $\pi_4(\Sigma K(A_4,1))$, is equal to $\Zset_4$; Kenzo's calculations have showed that the correct result (as later confirmed by the authors of \cite{MW10}) is $\Zset_{12}$. See \cite{RR12} for details on these calculations.


In a previous work \cite{RRS06}, we have proved that if $C$ is a chain complex with effective homology $C \lrdc D \rrdc EC$ and we define a filtration on $C$, then appropriate filtrations of $D$ and $EC$ can produce spectral sequences which are isomorphic to that of $C$. The different maps involved in the strong chain equivalence $C \lrrdc EC$ must satisfy some ``natural'' conditions with respect to the filtrations on $C$, $D$ and $EC$.

\begin{Defn}
Given two filtered chain complexes $C$ and $D$, a \emph{filtered chain complex morphism} $f: C \rightarrow D$ is a chain complex morphism which is compatible with the filtrations, that is to say,
$$
f(C^p_n) \subseteq D^p_n \mbox{ for each degree } n \mbox{ and filtration index } p.
$$
\end{Defn}

\begin{Defn}
Given two filtered complex morphisms $f,g: C \rightarrow D$ and a chain homotopy $h: f \simeq g$, we say that $h$ has \emph{order} $\leq s$ if
$$
h(C^p_n) \subseteq D^{p+s}_{n+1} \mbox{ for each degree } n \mbox{ and filtration index } p.
$$
\end{Defn}

The main result of our work in \cite{RRS06}, which will allow us to compute spectral sequences of complexes not of finite type, is the following theorem:

\begin{Thm}
\label{thm:spct-efhm}
Let $C$ be a chain complex with a filtration. Let us suppose that $C$ is an object with effective homology, such that there exists an equivalence $ C \stackrel{\rho_1}{\lrdc} D \stackrel{\rho_2}{\rrdc} EC$ with $\rho_1=(f_1,g_1,h_1)$ and $\rho_2=(f_2,g_2,h_2)$, and such that filtrations are also defined on the chain complexes $D$ and $EC$. If the maps $f_1$, $f_2$, $g_1$, and $g_2$ are  morphisms of filtered chain complexes and both homotopies $h_1$ and $h_2$ have order $\leq s$, then the spectral sequences of the complexes $C$ and $EC$ are isomorphic for $r>s$:
\[
E(C)^r_{p,q}\cong E(EC)^r_{p,q} \quad \mbox{for all } p,q \in \Zset \mbox{ and } r>s.
\]
\end{Thm}

This theorem makes it possible to compute spectral sequences of (complicated) filtered complexes with effective homology, obtaining in this way an algorithm which allows us to determine the different components of the spectral sequence. This algorithm follows the same idea as the effective homology method to determine homology groups of spaces, that is: given an effective chain complex $C$ with a filtration, the different components of the associated
spectral sequence can be computed by means of some elementary algorithms; on the other hand, if the filtered chain complex $C$ is not effective, but
with effective homology $C \lrdc D \rrdc
EC$, then the spectral sequence of $C$ can be determined thanks to the effective filtered chain complex $EC$.

Our algorithm has been implemented as a new module for Kenzo (the programs can be accessed at \cite{Rom06}), and allows us to determine the groups $E^r_{p,q}$ and the differential maps $d^r_{p,q}$ in all levels of the spectral sequence, and also the convergence level (that is, the stage $r$ for which the groups $E^r_{p,q}$ are already the final groups $E^\infty_{p,q}$ of the spectral sequence) for each dimension $n$ and the filtration of the homology groups induced by the filtration of the chain complex. The computations are done over the integer ring $R=\Zset$, and the groups $E^r_{p,q}$ (which are finitely generated Abelian groups) are given by a list of generators and a list of \emph{divisors} (the Betti number and the torsion coefficients of the group).

In particular, our programs can be applied to compute spectral sequences of bicomplexes and also the classical spectral sequences of Serre and Eilenberg-Moore (which are defined by means of filtered chain complexes), where complicated (not of finite type) spaces are involved. The details of the construction of the effective homology of the spaces which produce the Serre and Eilenberg-Moore spectral sequences can be found in \cite{RS06}. It was necessary to prove that the filtrations defined on the spaces involved in the effective homology of these complexes satisfy the necessary conditions of our Theorem \ref{thm:spct-efhm}. In the case of the Serre spectral sequence it was proved that the homotopy operators $h_1$ and $h_2$ have filtration order $\leq 1$, 
so that our algorithm makes it possible to determine the elements $E^r_{p,q}$ and $d^r_{p,q}$ of the Serre spectral sequence after $r \geq 2$. For the Eilenberg-Moore spectral sequence, the homotopy operators have order $\leq 0$ (that is, $h_1$ and $h_2$ do not increase the filtration index) and then our algorithm computes all stages of the spectral sequence. See \cite{RRS06} for details and some examples of calculations of these spectral sequences.

It is worth emphasizing here that the conditions about the filtrations and the maps in the effective homology of the initial complex required in Theorem~\ref{thm:spct-efhm} are necessary. The following example illustrates the fact that if these conditions are not satisfied, then the spectral sequence of the initial chain complex $C$ and that of the effective chain complex $EC$ are not isomorphic in all levels.

Let us consider a chain complex $C$ with only three non-null groups $C_0=\Zset$, $C_1=\Zset^2$ and $C_2=\Zset$, and differential maps $d_1$ and $d_2$ given respectively by the matrices $D_1=[1 \quad 0]$ and $D_2=\left[
                                                                                                                                                           \begin{array}{c}
                                                                                                                                                             0  \\
                                                                                                                                                             1  \\
                                                                                                                                                           \end{array}
                                                                                                                                                         \right]$. This chain complex is acyclic and it is not
difficult to find a reduction from $C$ to an (also effective) chain complex $EC$ which in this case has all components $EC_n=0$. The $h$ component in the reduction is given by $H_0=\left[
                                                                                                                                                           \begin{array}{c}
                                                                                                                                                             1  \\
                                                                                                                                                             0  \\
                                                                                                                                                           \end{array}
                                                                                                                                                         \right]$
and $H_1=[0 \quad 1]$. Let us call $a$ the generator of $C_0$, $b_1$ and $b_2$ the generators of $C_1$ and $c$ the generator of $C_2$. Then we filter the chain complex such that $C^1$ is the subcomplex generated by $a$ and $b_2$ and $C^2=C$. The level $1$ of the spectral sequence of the initial chain complex $C$ has groups $E^1_{1,-1}=\Zset$, $E^1_{1,0}=\Zset$, $E^1_{2,-1}=\Zset$ and $E^1_{2,0}=\Zset$, but the groups $E^r_{p,q}$ of the spectral sequence of the small chain complex $EC$ are null for every $r, p$ and $q$. In level $2$, the spectral sequence of $C$ is also null. One can observe that the homotopy operator $h$ in the reduction has filtration order $\leq 1$, so that Theorem \ref{thm:spct-efhm} claims that the spectral sequences of $C$ and $EC$ are isomorphic after level $2$, but we have seen that in level $1$ they are different.

\section{Computing persistent homology}
\label{sec:computing-prst}

In Section \ref{sec:spct-seq-infinite}, we have introduced a new module for the Kenzo system which computes the different components of the spectral sequence associated with a filtered chain complex, even in some cases where the chain complex has infinite type.

On the other hand, we have seen in Section \ref{sec:pers-hom-and-ss-relation} that given a filtered simplicial complex $K$, the associated spectral sequence and the persistent homology groups are related; more concretely, the homology classes of persistent index equal to $r$ correspond to the elements in the image of the differential
maps in the level $E^r$ of the spectral sequence, and a class which never dies corresponds to a
component in the final level $E^{\infty}$. It is clear then that if the different groups $E^r_{p,q}$ and the differential maps $d^r_{p,q}$ of the spectral sequence are known then we can also determine the groups $BD^{i,k}_n$ of elements which are born at $K^i$ and die entering $K^k$. Making use of the results of our programs for spectral sequences, we can then compute the images of the corresponding differential maps to determine also the groups $BD^{i,k}_n$ of a filtered chain complex. When working over a field, this information is sufficient to determine also the persistent homology groups $H^{i,j}_n$ and the filtration given by $H^{i,j,k}_n$; in the integer case, we have explained before that one can find extension problems in order to determine the groups $H^{i,j}_n$ of classes which are in $K^i$ and are still alive at $K^j$, and the \emph{intermediate} groups $H^{i,j,k}_n$.

In fact the computation of the groups $BD^{i,k}_n$ can be obtained by a small modification of our programs without doing the complete process of computing the corresponding groups and differential maps of the spectral sequence. Let us recall that a group $E^r_{p,q}$ in the spectral sequence is given by the formula:
\[
E^r_{p,q}=\frac{Z^r_{p,q}+ C^{p-1}_{p+q}}{d_{p+q+1}(Z^{r-1}_{p+r-1,q-r+2})+ C^{p-1}_{p+q}}
\]
and, as said before,
each class in $E^r_{p,q}$ is generated by an ``almost'' cycle of dimension $p+q$ (a chain whose boundary in~$K^p-K^{p-r}$ is empty but which may have non-empty boundary in $K^{p-r}$), and the elements of $E^r_{p,q}$ given by a real cycle $x$ (that is, $d (x)=0$), correspond to classes of $H_{p+q}(K^p)$ which are born at $K^p$ and are still alive at $K^{p+r-1}$, and then the persistence indexes of these classes are at least $r$.

It is not difficult to observe then that the groups $BD^{i,k}_n$ can be determined by the formula:
\[
BD_n^{i,k}=\frac{d_{n+1}(Z^{k-i}_{k,n-k+1}) + C^{i-1}_{n}}{d_{n+1}(Z^{k-i-1}_{k-1,n-k+2})+ C^{i-1}_{n}}
\]

If $K$ is a finite filtered simplicial complex, then our programs determine the different elements of the associated spectral sequence by means of some diagonalization algorithms on matrices. More concretely, the programs determine in particular the subgroups $Z^r_{p,q}$, $C^{p-1}_{p+q}$ and $d_{p+q+1}(Z^{r-1}_{p+r-1,q-r+2})$ which appear in the formula of Theorem \ref{thm:fltr-spsq} (which can be determined if $K$ is finite), and then calculate the desired quotient. The groups $BD_n^{i,k}$ are determined in terms of similar subgroups and then it has been very easy to adapt our programs in order to compute also $BD_n^{i,k}$ for finite (filtered) simplicial complexes.

One can observe that the groups $H^{i,j,k}_n$ can also be described as a quotient:
\[
H_n^{i,j,k}=\frac{(\kr d_n \cap C^{i-1}_n) + d_{n+1}(Z^{k-i}_{k,n-k+1})}{d_{n+1}(Z^{j-i}_{j,n-j+1})}=\frac{Z^{i-1}_{i-1,n-i+1} + d_{n+1}(Z^{k-i}_{k,n-k+1})}{d_{n+1}(Z^{j-i}_{j,n-j+1})}
\]

Again, the subgroups which appear in this new formula can be determined by our programs for spectral sequences so that one can compute directly the groups $H^{i,j,k}_n$ providing our double filtration for $H^j_n$.

 Finally, the \emph{total} persistent homology groups $H^{i,j}_n$ can also be expressed in terms of the subgroups involved in the spectral sequence, in this case:
\[
H_n^{i,j}=\frac{\kr d_n \cap C^{i}_{n}}{d_{n+1}(Z^{j-i}_{j,n-j+1})}  =\frac{Z^i_{i,n-i}}{d_{n+1}(Z^{j-i}_{j,n-j+1})}
\]
so that our programs can determine them. It is important to remark that this makes it possible to solve the possible extension problems that one could find when trying to deduce the groups $H^{i,j}_n$ and $H_n^{i,j,k}$ from $BD^{i,k}_n$.

In the infinite case, the effective homology method can be used to determine the groups $H^{i,j}_n$: let $C$ be a filtered chain complex with effective homology $C \lrdc D \rrdc EC$, then appropriate filtrations of $D$ and $EC$ can also produce the persistent homology groups. Again some ``natural'' conditions on the filtrations and the maps involved in the effective homology are necessary.

\begin{Thm}
\label{thm:prst-hom-efhm}
Let $C$ be a chain complex with a filtration. Let us suppose that $C$ is an object with effective homology, such that there exists an equivalence $ C \stackrel{\rho_1}{\lrdc} D \stackrel{\rho_2}{\rrdc} EC$ with $\rho_1=(f_1,g_1,h_1)$ and $\rho_2=(f_2,g_2,h_2)$, and such that filtrations are also defined on the chain complexes $D$ and $EC$. If the maps $f_1$, $f_2$, $g_1$, and $g_2$ are  morphisms of filtered chain complexes and both homotopies $h_1$ and $h_2$ have order $\leq s$, then the persistent homology groups $H^{i,j}_n$ of $C$ and $EC$ are (explicitly) isomorphic for $j-i\geq s$:
\[
H^{i,j}_n(C)\cong H^{i,j}_{n}(EC) \quad \mbox{for all } n \in \Nset \mbox{ and } j-i\geq s
\]
and the groups $H_n^{i,j,k}$ and $BD^{i,k}_n$ of $C$ and $EC$ are (explicitly) isomorphic for $k-i > s$:
$$
H_n^{i,j,k}(C)\cong H_n^{i,j,k}(EC) \quad \mbox{for all } n \in \Nset \mbox{ and } k-i > s $$
$$
BD^{i,k}_n(C)\cong BD^{i,k}_{n}(EC) \quad \mbox{for all } n \in \Nset \mbox{ and } k-i > s
$$
\end{Thm}

The proof of this theorem is not included here because is similar to that of Theorem \ref{thm:spct-efhm}, which can be found in \cite{RRS06}, and Proposition 3.5 in \cite[p. 331]{Mac63}. The isomorphisms between the corresponding groups are deduced from the compositions $f_2 g_1 : C \rightarrow EC$ and $f_1 g_2 : EC \rightarrow C$. In particular, if both homotopies $h_1$ and $h_2$ have order $0$ (that is, they are compatible with the filtration on $D$), then all groups $H^{i,j}_n$, $H_n^{i,j,k}$ and $BD^{i,k}_n$ of $C$ and $EC$ are isomorphic.

Let us observe that if $EC$ is an effective chain complex, then one can determine its persistent homology groups by means of elementary algorithms: each subcomplex $EC^i$ has finite type, so that its homology groups $H_n(EC^i)\equiv H^i_n$ are computable. Then the maps $f^{i,j}_n:H^{i}_n \rightarrow H^{j}_n$ can be expressed by means of finite matrices and therefore we can compute the groups $H^{i,j}_n=\im f^{i,j}_n$. Similarly, the groups $H^{i,j,k}_n=H^{i,j}_n \cap (f_n^{j,k})^{-1}(H^{i-1,k}_n) \subseteq H^{i,j}_n \subseteq H^j_n$ and $BD_n^{i,k}={{H}^{i,i,k}_n}/{{H}^{i,i,k-1}_n}$ of $EC$ can be computed by means of matrix diagonalization. Thanks to Theorem \ref{thm:prst-hom-efhm}, we can also compute the persistent homology groups of the initial (big) chain complex $C$ by means of those of~$EC$. The following algorithm is therefore obtained.

\newpage
\parskip1ex
\begin{Alg}
\label{alg:prst-hom}
\inp
\algitem
\item a filtered chain complex $C$ with effective homology $ C \stackrel{\rho_1}{\lrdc} D \stackrel{\rho_2}{\rrdc} EC$ with $\rho_1=(f_1,g_1,h_1)$ and $\rho_2=(f_2,g_2,h_2)$, and such that filtrations are also defined on $D$ and $EC$ and such that the maps $f_1$, $f_2$, $g_1$, and $g_2$ are  morphisms of filtered chain complexes and both homotopies $h_1$ and $h_2$ have order $\leq s$,
\item the numbers $n,i,j \in \Nset$ such that $j-i\geq s$.
\end{list}
\outp a basis-divisors description of the persistent homology group $H^{i,j}_n(C)$, in other words,
\algitem
\item a list of combinations $(c_1, \ldots , c_{t+\alpha})$ which generate the group,
\item and a list of non-negative integers $(\beta_1, \ldots, \beta_t,0,\stackrel{\alpha}{\ldots},0)$ (such that $\beta_i$ divides $\beta_{i+1}$) where $\alpha$ is the Betti number of the group and $\beta_1,\ldots,\beta_t$ are the torsion coefficients.
\end{list}
\end{Alg}

Similar algorithms can also be constructed computing a basis-divisors description of the groups $H^{i,j,k}_n(C)$ and $BD^{i,k}_n(C)$, in these cases for $k-i > s$. These algorithms have been implemented in Common Lisp enhancing Kenzo and making use of our module for computing spectral sequences of filtered complexes. The code of our programs can be found in~\cite{Rom12}.

Again the conditions supposed for the maps in the effective homology and the filtrations are necessary; considering the example of the chain complex $C$ introduced at the end of Section \ref{sec:spct-seq-infinite} with three non-null groups $C_0=\Zset$, $C_1=\Zset^2$ and $C_2=\Zset$, and differential maps $d_1\equiv [1 \quad 0]$ and $d_2\equiv \left[
                                                                                                                                                           \begin{array}{c}
                                                                                                                                                             0  \\
                                                                                                                                                             1  \\
                                                                                                                                                           \end{array}
                                                                                                                                                         \right]$, which can be reduced to the null chain
 complex, one can observe that there are some homology groups of persistence index $1$ for the \emph{big} chain complex but those of the small one are null.

Once a strong chain equivalence is established between an initial chain complex $C$ and an effective chain complex $EC$, we can obtain the generators of the homology groups of $C$ expressed as cycles on $C$. To this aim, we get the (representatives of) generators of the homology groups of $EC$ as cycles in $EC$ (it is a by-product of the diagonalization process to determine Betti numbers and torsion coefficients), and then we apply on them the composition $f_1 g_2$ of the chain equivalence $\varepsilon$, getting the announced cycles over $C$. In fact, the chain equivalence produces a complete solution of the \emph{homological problem} for $C$; see the statement of this problem in \cite{Ser09}. If the chain complexes $C$ and $EC$ are filtered and the equivalence $\varepsilon$ satisfies the hypothesis of Theorem \ref{thm:prst-hom-efhm}, the same process as before produces the generators of the persistent homology groups $H^{i,j}_n(C)$, $H^{i,j,k}_n(C)$  and $BD^{i,k}_n(C)$. This opens the possibility of a \emph{qualitative} study of persistent homology, going beyond the traditional \emph{quantitative} analysis (based, for instance, on barcodes). With our approach we can trace the born and death moments of particular cycles, and their contribution to the persistent homology groups.

 When trying to compute the persistent homology groups $H^{i,j}_n$, $H^{i,j,k}_n$ and $BD^{i,k}_n$  of a chain complex with effective homology, it will be necessary to determine the order of the homotopy operators in the chain equivalence. As already said in Section~\ref{sec:spct-seq-infinite}, in the case of the Serre spectral sequence those homotopy operators have filtration order $\leq 1$, 
so that our algorithms make it possible to determine the elements $H^{i,j}_n$  for $j-i \geq 1$, and $BD^{i,k}_n$ and $H^{i,j,k}_n$ for $k-i>1$, which correspond to intervals of length greater than $1$. For the Eilenberg-Moore spectral sequence, the homotopy operators have order $\leq 0$ and then our algorithms compute all persistent homology groups.

\section{Examples and applications}
\label{sec:examples_and_applications}

\subsection{Integer persistence of torus and Klein bottle}

\begin{figure}
\begin{center}
\begin{tikzpicture}
\draw (-1,0) node{\includegraphics[scale=0.5]{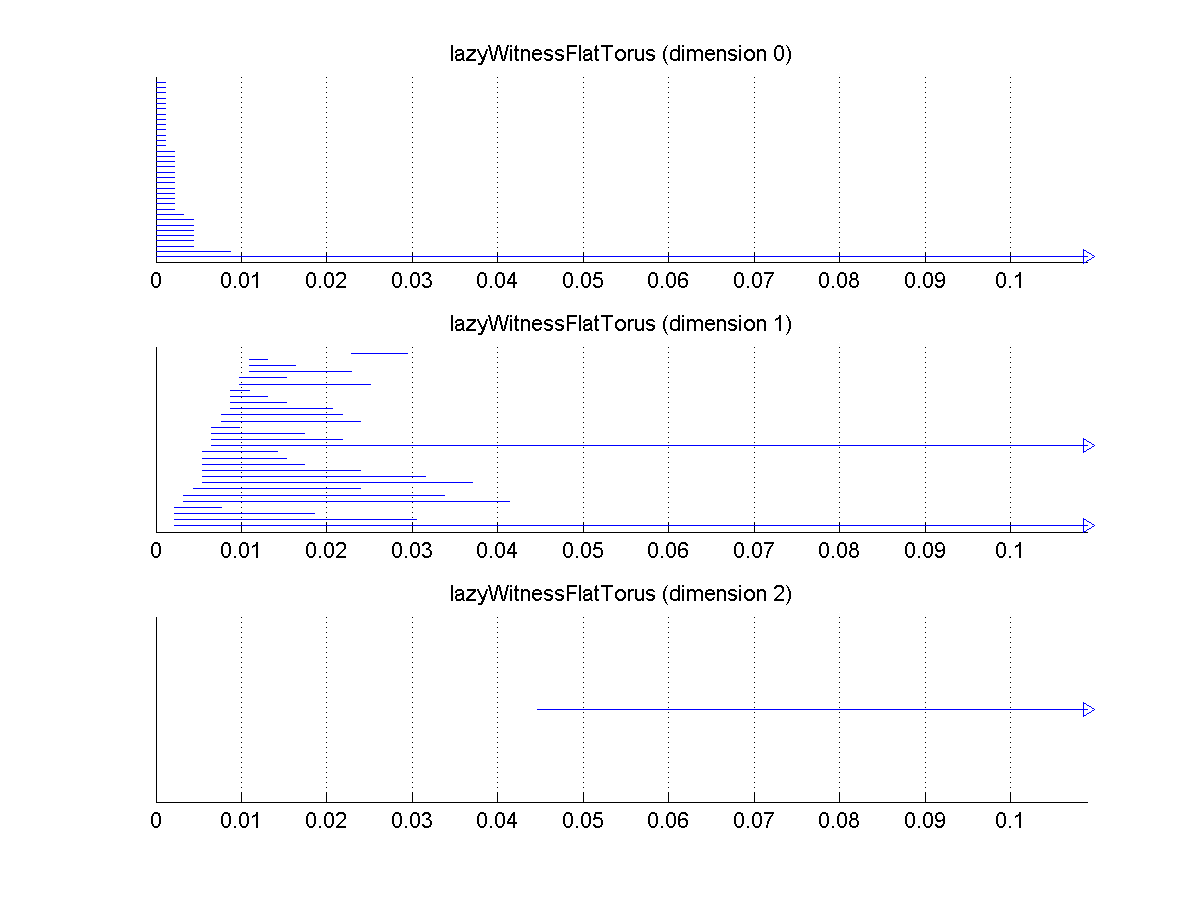}};
\end{tikzpicture}
\end{center}
\caption{Torus barcode diagram in $\Zset_2$.}
\label{fig:barcode_torus_z2}
\end{figure}

\begin{figure}
\begin{center}
\begin{tikzpicture}
\draw (-1,0) node{\includegraphics[scale=0.5]{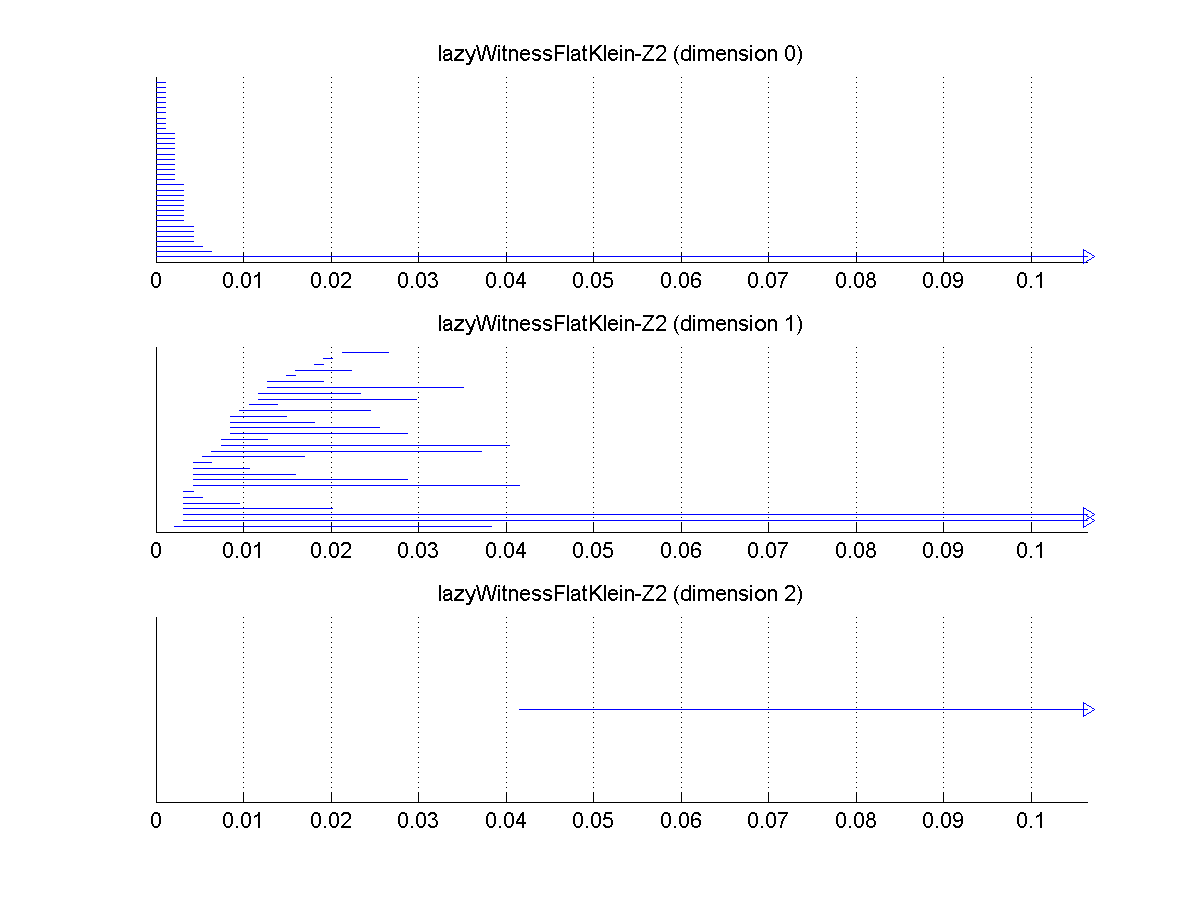}};
\end{tikzpicture}
\end{center}
\caption{Klein bottle barcode diagram in $\Zset_2$.}
\label{fig:barcode_klein_z2}
\end{figure}

Let us make use of our programs for computing the persistent homology groups with integer coefficients of a torus and a Klein bottle. To this aim, we use the program \emph{JavaPlex} \cite{javaPlex} to create an explicit metric space for $1000$ random points on a flat torus (respectively on a flat Klein bottle) and then construct an associated \emph{lazy witness} filtered simplicial complex \cite{SC04}. Computing with JavaPlex the persistent homology groups of both spaces with coefficients on $\Zset_2$, we obtain similar barcode diagrams, see Figures \ref{fig:barcode_torus_z2} and \ref{fig:barcode_klein_z2}. In particular, in dimension $1$ one can observe in both cases two long intervals; from these results it is not possible to distinguish if each one of the complexes corresponds to a torus or a Klein bottle.

The two previous filtered simplicial complexes (associated respectively with the torus and the Klein bottle) have been imported in Kenzo by means of OpenMath \cite{OM}, an XML standard for representing mathematical objects. After extending the OpenMath libraries to represent filtered simplicial complexes, we have written the corresponding \emph{parsers} to export and import  OpenMath code from and into JavaPlex and Kenzo respectively. In this way, the filtered simplicial complexes of the torus and the Klein bottle constructed in JavaPlex are translated to OpenMath code; this code is stored in a file and this file is read by Kenzo, which constructs the corresponding filtered simplicial complexes. Once the complexes are implemented in Kenzo, our new functions computing persistent homology with integer coefficients can be applied.

For example, the simplicial complex associated with the Klein bottle is stored in Kenzo in a variable called \texttt{kleinb}. Then the new function \texttt{prst-hmlg-group} can be applied; it has as inputs a filtered chain complex $C$ and integers $i, k$, and $n$ and computes the group $BD^{i,k}_n(C)$ with the homology classes that are born at $C^i$ and die entering $C^k$. For example, the persistent homology group $BD^{0,3}_0$ is equal to $\Zset^8$}:

{\footnotesize
\begin{verbatim}
> (prst-hmlg-group kleinb 0 3 0)
Persistent Homology BD^{0,3}_0
Component Z
Component Z
Component Z
Component Z
Component Z
Component Z
Component Z
Component Z
\end{verbatim}}

\normalsize Similarly for the torus, stored in the variable \texttt{torus}, we obtain for instance $BD^{0,4}_0=\Zset^6$:

 {\footnotesize
\begin{verbatim}
> (prst-hmlg-group torus 0 4 0)
Persistent Homology BD^{0,4}_0
Component Z
Component Z
Component Z
Component Z
Component Z
Component Z
\end{verbatim}}

\normalsize
In dimension $1$, barcode diagrams of Figures \ref{fig:barcode_torus_z2} and \ref{fig:barcode_klein_z2} show that both the Klein bottle and the torus have two long intervals (corresponding to homology classes which do not die and persist forever). However, using our programs for computing persistent homology with $\Zset$-coefficients one can see that the corresponding groups are different and this makes it possible to distinguish the Klein bottle and the torus. In this case, we use the new function \texttt{total-prst-hmlg-group} for computing the groups $H^{i,j}_n$.

{\footnotesize
\begin{verbatim}
> (total-prst-hmlg-group torus 10 99 1)
Persistent Homology H^{10,99}_1
Component Z
Component Z
> (total-prst-hmlg-group kleinb 10 99 1)
Persistent Homology H^{10,99}_1
Component Z/2Z
Component Z
\end{verbatim}}

\normalsize Moreover, one can detect that the component $\Zset_2$ appears at stage $39$ of the filtration:

{\footnotesize
\begin{verbatim}
> (total-prst-hmlg-group kleinb 3 38 1)
Persistent Homology H^{3,38}_1
Component Z
Component Z
> (total-prst-hmlg-group kleinb 3 39 1)
Persistent Homology H^{3,39}_1
Component Z/2Z
Component Z
> (prst-hmlg-group kleinb 3 39 1)
Persistent Homology BD^{3,39}_1
Component Z
\end{verbatim}}

\normalsize

\subsection{Computing persistence in Postnikov towers}
\label{subsec:postnikov}

We consider now the space $X_3$ of a Postnikov tower \cite{May67} with $\pi_i=\Zset_2$ at each stage and the ``simplest'' non-trivial Postnikov invariant. The theoretical details of the construction of this space are not included here, they can be found in \cite{RS05b}. This complex can be built by Kenzo by means of the following statements:

{\footnotesize
\begin{verbatim}
> (setf X2 (k-z2 2))
[K13 Abelian-Simplicial-Group]
> (setf k3 (chml-clss X2 4))
[K125 Cohomology-Class on K30 of degree 4]
> (setf tau3 (z2-whitehead X2 k3))
[K140 Fibration K13 -> K126]
> (setf X3 (fibration-total tau3))
[K146 Kan-Simplicial-Set]
\end{verbatim}}

\normalsize
The result is a Kan simplicial set, stored in the variable \texttt{X3}. The effective homology of $X_3$ is directly built by Kenzo (see \cite{RS06} for details on the construction of the chain equivalence) and is stored in the slot
\texttt{efhm}. The effective chain complex can be accessed as follows:

{\footnotesize
\begin{verbatim}
> (setf effX3 (rbcc (efhm X3)))
[K344 Chain-Complex]
\end{verbatim}}

\normalsize
 The space $X_3$ is a twisted Cartesian product $X_3=K(\Zset_2,3)\times_{k_3} K(\Zset_2,2)$, total space of a fibration $K(\Zset_2,3) \hookrightarrow X_3 \rightarrow K(\Zset_2,2)$. 
 The object $\texttt{X3}$ is already of finite type, but its effective homology gives us an associated effective chain complex which is much smaller. For instance, \texttt{X3} has $1,043,600$ generators in dimension $5$ and the small chain complex \texttt{effX3} has only $6$.

One can compute the Serre spectral sequence associated with \texttt{X3}. For this task it is necessary to define filtrations for both spaces \texttt{X3} and \texttt{effX3}. Following the \emph{classical} filtrations defined on cartesian and tensor products defined by Serre, the filtration starts at stage $0$, that is, the first (non-null) subcomplex is $K^0$ and not $K^1$ as considered in the previous sections of this paper.

{\footnotesize\begin{verbatim}
> (change-chcm-to-flcc X3 crpr-flin '(crpr-flin))
[K146 Filtered-Kan-Simplicial-Set]
> (change-chcm-to-flcc effX3 tnpr-flin '(tnpr-flin))
[K344 Filtered-Chain-Complex]
\end{verbatim}}

\begin{figure}
\footnotesize
\mbox{
 \mbox{\begin{xy}<0.7cm,0cm>:<0cm,0.7cm>::
 (0,0)*{\Zset} ; (1,0)*{0} ; (2,0)*{\Zset_2} ; (3,0)*{0} ; (4,0)*{\Zset_4} ; (5,0)*{\Zset_2} ; (6,0)*{\Zset_2} ; (7,0)*{\Zset_2} ;
 (0,1)*{0} ; (1,1)*{0} ; (2,1)*{0} ; (3,1)*{0} ; (4,1)*{0} ; (5,1)*{0} ; (6,1)*{0} ;
 (0,2)*{0} ; (1,2)*{0} ; (2,2)*{0} ; (3,2)*{0} ; (4,2)*{0} ; (5,2)*{0} ;
  (0,3)*{\Zset_2} ; (1,3)*{0} ; (2,3)*{\Zset_2} ; (3,3)*{\Zset_2} ; (4,3)*{\Zset_2} ;
  (0,4)*{0} ; (1,4)*{0} ; (2,4)*{0} ; (3,4)*{0} ;
  (0,5)*{\Zset_2} ; (1,5)*{0} ; (2,5)*{\Zset_2} ;
 (0,6)*{\Zset_2} ; (1,6)*{0} ;
(0, 7)*{\Zset_2} ;
(7.5,7)*{\fbox{$r=4$}} ;
 (7.7,0.1)* !D{p} ; (0.1,7.7) * !L{q} ;

 \ar@{.>} (0,0);(8,0)
 \ar@{.>}  (0,0);(0,8)

\ar|(0.5){\times 1} (3.8,0.2) ; (0.2,2.8)
\ar|(0.5){\cong} (5.8,0.2) ; (2.2,2.8)
\ar|(0.5){0} (3.8,3.2) ; (0.2,5.8)
\ar|(0.5){0} (6.8,0.2) ; (3.2,2.8)
\end{xy}}
\hspace{0.5cm}
\mbox{\begin{xy}<0.7cm,0cm>:<0cm,0.7cm>::
 (0,0)*{\Zset} ; (1,0)*{0} ; (2,0)*{\Zset_2} ; (3,0)*{0} ; (4,0)*{\Zset_2} ; (5,0)*{\Zset_2} ; (6,0)*{0} ; (7,0)*{\Zset_2} ;
 (0,1)*{0} ; (1,1)*{0} ; (2,1)*{0} ; (3,1)*{0} ; (4,1)*{0} ; (5,1)*{0} ; (6,1)*{0} ;
 (0,2)*{0} ; (1,2)*{0} ; (2,2)*{0} ; (3,2)*{0} ; (4,2)*{0} ; (5,2)*{0} ;
  (0,3)*{0} ; (1,3)*{0} ; (2,3)*{0} ; (3,3)*{\Zset_2} ; (4,3)*{0} ;
  (0,4)*{0} ; (1,4)*{0} ; (2,4)*{0} ; (3,4)*{0} ;
  (0,5)*{\Zset_2} ; (1,5)*{0} ; (2,5)*{\Zset_2} ;
 (0,6)*{\Zset_2} ; (1,6)*{0} ;
(0, 7)*{\Zset_2} ;
(7.5,7)*{\fbox{$r=5$}} ;
 (7.7,0.1)* !D{p} ; (0.1,7.7) * !L{q} ;

 \ar@{.>} (0,0);(8,0)
 \ar@{.>}  (0,0);(0,8)

\end{xy}}}
\normalsize

\caption{levels $4$ and $5$ of the Serre spectral sequence of $X_3=K(\Zset_2,3)\times_{k_3} K(\Zset_2,2)$.}\label{fig:serre-spctseq}
\end{figure}
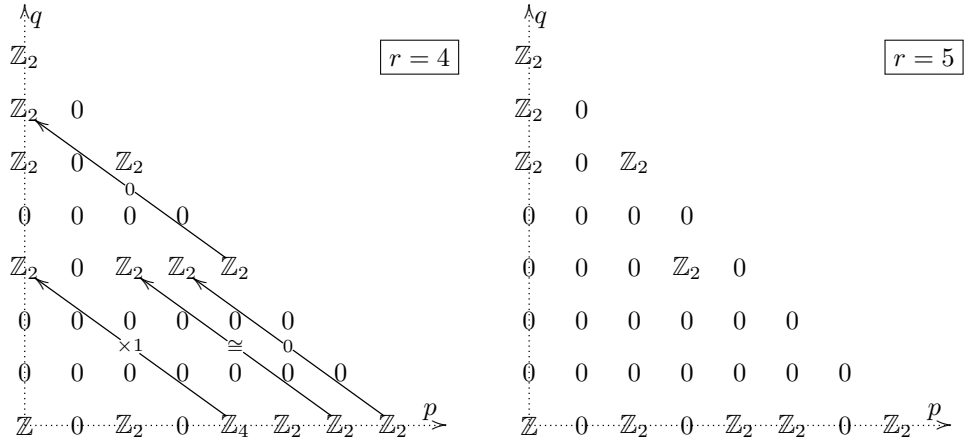

\normalsize Figure \ref{fig:serre-spctseq} shows the groups $E^r_{p,q}$ of the Serre spectral sequence, obtained thanks to our programs. The two diagrams correspond to the \emph{critical} levels $r=4$ and $r=5$ of the spectral sequence (only the groups $E^r_{p,q}$ with $p+q < 8$ are drawn). The groups $E^4_{p,q}$ are the same as the corresponding $E^2_{p,q}$ and $E^3_{p,q}$, which means that the first non-null differential maps appear at stage $r=4$. For $p+q\leq 6$, the spectral sequence converges at level $r=5$, that is to say, $E^5_{p,q}=E^\infty_{p,q}$. For $p+q=7$, the convergence is reached at the stage $r=9$; the groups $E^5_{0,7}\cong \Zset_2$ and $E^5_{2,5}\cong \Zset_2$ die at levels $9$ and $7$ respectively.

Let us compute now the persistent homology groups of the filtered space~$X_3$ making use of the effective homology. In this case, the results will only be valid for persistence index $k-i> 1$. Let us compute for example the group $BD^{0,4}_3$:

{\footnotesize\begin{verbatim}
> (prst-hmlg-group X3 0 4 3)
Persistent Homology BD^{0,4}_3
Component Z/2Z
\end{verbatim}

\normalsize This means that in dimension $3$ there is a persistent class which is born at $K^0$, is still alive at $K^3$ but dies entering $K^4$. One can observe this corresponds to the image of the differential $d^4_{4,0}: E^4_{4,0} \rightarrow E^4_{0,3}$ in the spectral sequence.

In dimension $7$, the groups $BD^{0,8}_7$, $BD^{2,8}_7$ and $BD^{4,8}_7$ are equal to $\Zset_2$, that is, there are three classes which are born respectively at $K^0$, $K^2$ and $K^4$ and die entering $K^8$).

\newpage
{\footnotesize\begin{verbatim}
> (prst-hmlg-group X3 0 8 7)
Persistent Homology BD^{0,8}_7
Component Z/2Z
> (prst-hmlg-group X3 2 8 7)
Persistent Homology BD^{2,8}_7
Component Z/2Z
> (prst-hmlg-group X3 4 8 7)
Persistent Homology BD^{4,8}_7
Component Z/2Z
\end{verbatim}

\normalsize Let us observe that in this case these intervals $BD^{i,k}_7$ do not determine the \emph{total} persistent homology groups $H^{i,j}_7$ because several extensions are possible. However, the groups $H^{i,j}_n$ can be computed with our new function \texttt{total-prst-hmlg-group}, which allows us to solve the extension problems. For example, the group $H^{4,7}_7$ is equal to $\Zset_2 \oplus \Zset_4$:

{\footnotesize\begin{verbatim}
> (total-prst-hmlg-group X3 4 7 7)
Persistent Homology H^{4,7}_7
Component Z/2Z
Component Z/4Z
\end{verbatim}

\normalsize Although we do not include here the examples of computations, it is also possible to compute with our programs the generators of the different groups. In this particular case it is interesting to say that the generators obtained for $H^{4,7}_7=\Zset_2 \oplus \Zset_4$ are not any of the generators of the groups $BD^{0,8}_7$, $BD^{2,8}_7$ and $BD^{4,8}_7$.

All the persistent groups computed by our programs can be described by means of the integer barcode diagram of Figure \ref{fig:serre-barcode}. Let us observe that there are two non-trivial extensions (the one explained for $H_7(K^4)$ and another one in $H_5(K^5)$). Figure \ref{fig:serre-barcode-2} includes the alternative description of the barcode diagram where the non-trivial extensions are  represented by joining the corresponding intervals. 

\begin{figure}
\centering
 \begin{tikzpicture}
  \draw \foreach \s in {0,...,8} {(-3.45+\s,.7) node {{\small $\s$}}};
  \draw[step=1,gray!20!white,very thin,fill=white] \foreach \s in {0,...,8} {(-3.45+\s,.45) -- (-3.45+\s,-9.5)};

 \draw (-5.7,-0.2) node {$H_0$};
 \draw (0,-0.2) node{
 \begin{tikzpicture}
 \draw (-.4,0.2) -- (-.6,0.2) -- (-.6,-.2) -- (-.4,-.2);
 \draw [*->] (1.25,0) -- (9.85,0);
 \draw (0.25,0) node{$\mathbb{Z} \quad \mathbb{Z}$};
 \draw (10,0.2) -- (10.2,.2) -- (10.2,-.2) -- (10,-.2);
 \end{tikzpicture}};

 \draw (-5.7,-.95) node {$H_1$};
 \draw (0,-.95) node{
 \begin{tikzpicture}
 \draw (-.4,0.2) -- (-.6,0.2) -- (-.6,-.2) -- (-.4,-.2);
 \draw (10,0.2) -- (10.2,.2) -- (10.2,-.2) -- (10,-.2);
 \end{tikzpicture}};

 \draw (-5.7,-1.7) node {$H_2$};
 \draw (0,-1.7) node{
 \begin{tikzpicture}
 \draw (-.4,0.2) -- (-.6,0.2) -- (-.6,-.2) -- (-.4,-.2);
    \draw (0.25,0) node{$\mathbb{Z}_2 \ \mathbb{Z}_2$};
 \draw [*->] (3.25,0) -- (9.85,0);
 \draw (10,0.2) -- (10.2,.2) -- (10.2,-.2) -- (10,-.2);
 \end{tikzpicture}};

 \draw (-5.7,-2.45) node {$H_3$};
 \draw (0,-2.45) node{
 \begin{tikzpicture}
 \draw (-.4,0.2) -- (-.6,0.2) -- (-.6,-.2) -- (-.4,-.2);
 \draw [*-o] (1.25,0) -- (5.45,0);
 \draw (0.25,0) node{$\mathbb{Z}_2 \ \mathbb{Z}_2$};
 \draw (10,0.2) -- (10.2,.2) -- (10.2,-.2) -- (10,-.2);
 \end{tikzpicture}};

 \draw (-5.7,-3.2) node {$H_4$};
 \draw (0,-3.2) node{
 \begin{tikzpicture}
 \draw (-.4,0.2) -- (-.6,0.2) -- (-.6,-.2) -- (-.4,-.2);
 \draw [*->] (5.25,0) -- (9.85,0);
     \draw (0.25,0) node{$\mathbb{Z}_2 \ \mathbb{Z}_2$};
 \draw (10,0.2) -- (10.2,.2) -- (10.2,-.2) -- (10,-.2);
 \end{tikzpicture}};

 \draw (-6.1,-4.6) node {$H_5$};
 \draw (0,-4.6) node{
 \begin{tikzpicture}
 \draw (-1.2,0.7) -- (-1.4,0.7) -- (-1.4,-0.9) -- (-1.2,-0.9);
 \draw [*->] (0.835,0.5) -- (9.85,0.5);
 \draw (-0.35,0.5) node{$\quad \mathbb{Z}_2 \ \ \quad \mathbb{Z}_2$};
   \draw [*-o] (2.835,-0.1) -- (7,-0.1);
 \draw (-0.35,-0.1) node{$\mathbb{Z}_2\oplus \mathbb{Z}_2 \ \mathbb{Z}_2$};
  \draw [*->] (5.83,-0.7) -- (9.85,-0.7);
  \draw (-0.35,-0.7) node{$\mathbb{Z}_4\oplus \mathbb{Z}_2 \ \mathbb{Z}_2$};
  \draw (10,0.7) -- (10.2,.7) -- (10.2,-0.9) -- (10,-0.9);
 \end{tikzpicture}};

 \draw (-6.1,-6.25) node {$H_6$};
 \draw (0,-6.25) node{
 \begin{tikzpicture}
 \draw (-1.2,0.7) -- (-1.4,0.7) -- (-1.4,-.3) -- (-1.2,-.3);
 \draw [*->] (0.835,0.5) -- (9.85,0.5);
 \draw (-0.35,0.5) node{$\quad \mathbb{Z}_2 \ \ \quad \mathbb{Z}_2$};
   \draw [*->] (2.835,-0.1) -- (9.85,-0.1);
 \draw (-0.35,-0.1) node{$\mathbb{Z}_2\oplus \mathbb{Z}_2 \ \mathbb{Z}_2$};
 \draw (10,0.7) -- (10.2,.7) -- (10.2,-.3) -- (10,-.3);
 \end{tikzpicture}};

 \draw (-6.1,-8.2) node {$H_7$};
 \draw (0,-8.2) node{
 \begin{tikzpicture}
 \draw (-1.2,0.7) -- (-1.4,0.7) -- (-1.4,-1.5) -- (-1.2,-1.5);
 \draw [*-o] (0.835,0.5) -- (9,0.5);
 \draw (-0.35,0.5) node{$\quad \mathbb{Z}_2 \ \ \quad \mathbb{Z}_2$};
   \draw [*-o] (2.835,-0.1) -- (9,-0.1);
 \draw (-0.35,-0.1) node{$\mathbb{Z}_2\oplus \mathbb{Z}_2 \ \mathbb{Z}_2$};
    \draw [*-o] (4.835,-0.7) -- (9,-0.7);
 \draw (-0.35,-0.7) node{$\mathbb{Z}_4\oplus \mathbb{Z}_2 \ \mathbb{Z}_2$};
 \draw [*->] (7.835,-1.3) -- (9.85,-1.3);
 \draw (-0.35,-1.3) node{$\mathbb{Z}_4\oplus \mathbb{Z}_2^2 \ \mathbb{Z}_2$};

 \draw (10,0.7) -- (10.2,.7) -- (10.2,-1.5) -- (10,-1.5);
 \end{tikzpicture}};

\end{tikzpicture}
\caption{integer barcode diagram of the space $X_3=K(\Zset_2,3)\times_{k_3} K(\Zset_2,2)$.}\label{fig:serre-barcode}
\end{figure}
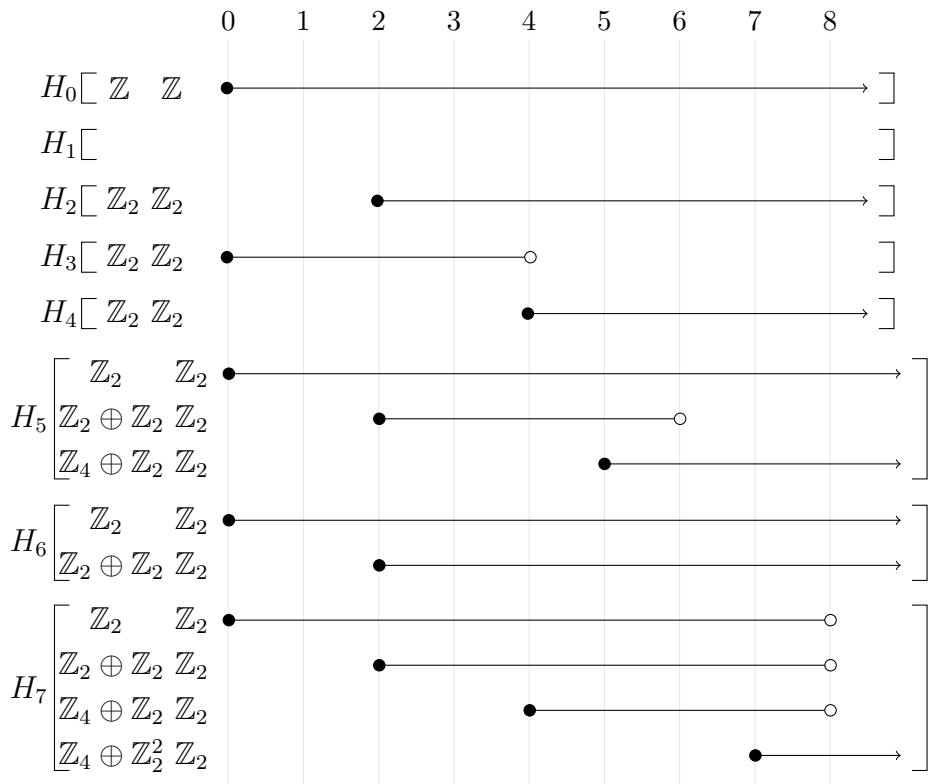

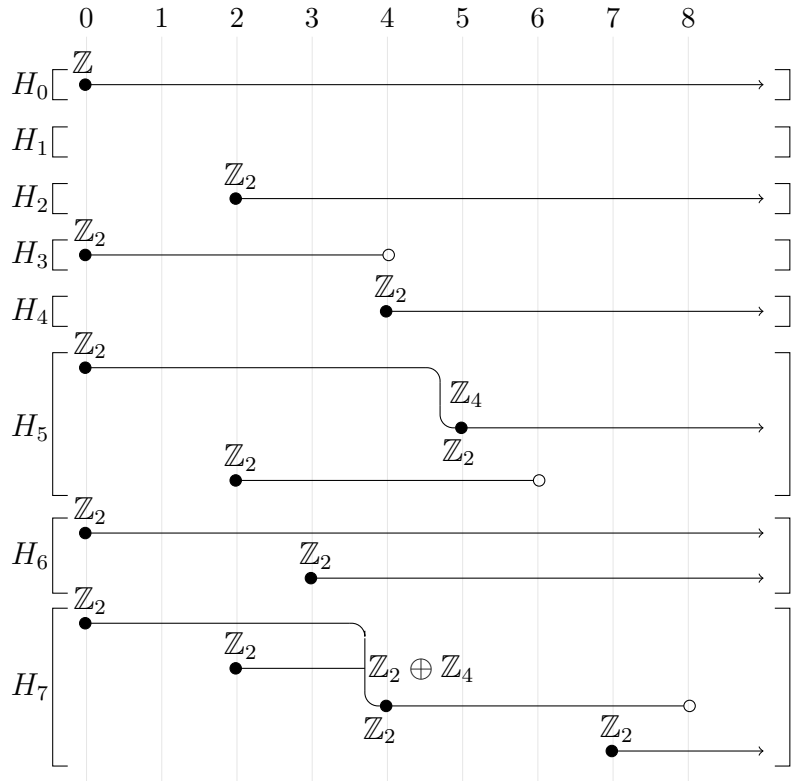
\begin{figure}
\centering
 \begin{tikzpicture}

  \draw \foreach \s in {0,...,8} {(-4.45+\s,.7) node {{\small $\s$}}};
  \draw[step=1,gray!20!white,very thin,fill=white] \foreach \s in {0,...,8} {(-4.45+\s,.45) -- (-4.45+\s,-9.5)};

 \draw (-5.2,-0.2) node {$H_0$};
 \draw (0,0) node{
 \begin{tikzpicture}
 \draw (-.4,0.2) -- (-.6,0.2) -- (-.6,-.2) -- (-.4,-.2);
  \draw (-0.2,0.3) node {$\Zset$};
 \draw [*->] (-.25,0) -- (8.85,0);
 \draw (9,0.2) -- (9.2,.2) -- (9.2,-.2) -- (9,-.2);
 \end{tikzpicture}};

 \draw (-5.2,-.95) node {$H_1$};
 \draw (0,-.95) node{
 \begin{tikzpicture}
 \draw (-.4,0.2) -- (-.6,0.2) -- (-.6,-.2) -- (-.4,-.2);
 \draw (9,0.2) -- (9.2,.2) -- (9.2,-.2) -- (9,-.2);
 \end{tikzpicture}};

 \draw (-5.2,-1.7) node {$H_2$};
 \draw (0,-1.5) node{
 \begin{tikzpicture}
 \draw (-.4,0.2) -- (-.6,0.2) -- (-.6,-.2) -- (-.4,-.2);
   \draw (1.9,0.3) node {$\Zset_2$};
 \draw [*->] (1.75,0) -- (8.85,0);
 \draw (9,0.2) -- (9.2,.2) -- (9.2,-.2) -- (9,-.2);
 \end{tikzpicture}};

 \draw (-5.2,-2.45) node {$H_3$};
 \draw (0,-2.25) node{
 \begin{tikzpicture}
 \draw (-.4,0.2) -- (-.6,0.2) -- (-.6,-.2) -- (-.4,-.2);
 \draw [*-o] (-.25,0) -- (3.95,0);
    \draw (-0.1,0.3) node {$\Zset_2$};
 \draw (9,0.2) -- (9.2,.2) -- (9.2,-.2) -- (9,-.2);
 \end{tikzpicture}};

 \draw (-5.2,-3.2) node {$H_4$};
 \draw (0,-3) node{
 \begin{tikzpicture}
 \draw (-.4,0.2) -- (-.6,0.2) -- (-.6,-.2) -- (-.4,-.2);
 \draw [*->] (3.75,0) -- (8.85,0);
    \draw (3.9,0.3) node {$\Zset_2$};
 \draw (9,0.2) -- (9.2,.2) -- (9.2,-.2) -- (9,-.2);
 \end{tikzpicture}};

 \draw (-5.2,-4.75) node {$H_5$};
 \draw (0,-4.5) node{
 \begin{tikzpicture}
 \draw (-.4,0.7) -- (-.6,0.7) -- (-.6,-1.2) -- (-.4,-1.2);
 \draw [*-] (-.25,0.5) -- (4.35,0.5);
 \draw (-.1,0.8) node {$\Zset_2$};
  \draw [*->] (4.75,-0.3) -- (8.85,-0.3);
  \draw (4.8,-0.3) node[anchor=north]{$\mathbb{Z}_2$};
  \draw[rounded corners=5pt] (4.35,0.5) -- (4.55,0.5) --  (4.55,0.3);
  \draw (4.9,0.15) node{$\mathbb{Z}_4$};
\draw (4.55,0.3) -- (4.55,0);
\draw[rounded corners=5pt] (4.55,0) -- (4.55,-0.3) --  (4.75,-0.3);

  \draw [*-o] (1.75,-1) -- (5.95,-1);
        \draw (1.9,-0.7) node {$\Zset_2$};
 \draw (9,0.7) -- (9.2,.7) -- (9.2,-1.2) -- (9,-1.2);
 \end{tikzpicture}};

 \draw (-5.2,-6.45) node {$H_6$};
 \draw (0,-6.25) node{
 \begin{tikzpicture}
 \draw (-.4,0.7) -- (-.6,0.7) -- (-.6,-.3) -- (-.4,-.3);
 \draw [*->] (-.25,0.5) -- (8.85,0.5);
      \draw (-0.1,0.8) node {$\Zset_2$};
 \draw [*->] (2.75,-0.1) -- (8.85,-0.1);
      \draw (2.9,0.2) node {$\Zset_2$};
 \draw (9,0.7) -- (9.2,.7) -- (9.2,-.3) -- (9,-.3);
 \end{tikzpicture}};

 \draw (-5.2,-8.2) node {$H_7$};
 \draw (0,-8) node{
 \begin{tikzpicture}
 \draw (-.4,0.7) -- (-.6,0.7) -- (-.6,-1.4) -- (-.4,-1.4);
 \draw [*-] (-.25,0.5) -- (3.35,0.5);
       \draw (-0.1,0.8) node {$\Zset_2$};
 \draw [*-] (1.75,-0.1) -- (3.55,-0.1);
       \draw (1.9,0.2) node {$\Zset_2$};
 \draw [*-o] (3.75,-.6) -- (7.95,-.6);
       \draw (3.75,-.6) node[anchor=north] {$\Zset_2$};
  \draw[rounded corners=5pt] (3.35,0.5) -- (3.55,0.5) --  (3.55,0.4);
\draw (3.55,0.3) -- (3.55,-0.4);
\draw[rounded corners=5pt] (3.55,-0.4) -- (3.55,-0.6) --  (3.75,-0.6);
  \draw (4.3,-0.1) node{$\mathbb{Z}_2 \oplus \mathbb{Z}_4$};
 \draw [*->] (6.75,-1.2) -- (8.85,-1.2);
       \draw (6.9,-.9) node {$\Zset_2$};
 \draw (9,0.7) -- (9.2,.7) -- (9.2,-1.4) -- (9,-1.4);
 \end{tikzpicture}};

\end{tikzpicture}
\caption{alternative description of the integer barcode diagram of
\mbox{$X_3=K(\Zset_2,3)\times_{k_3} K(\Zset_2,2)$.}}\label{fig:serre-barcode-2}
\end{figure}

\section{Conclusions}

It is a general principle that a computational perspective can
shed new light on theoretical mathematical concepts and results.
In this paper, we have particularized this principle in the case
of the relationship between persistent homology and spectral
sequences. It is quite evident that this relationship exists
(simply observing that both notions can be defined from a same
object: a simplicial filtration), and several papers and authors
remarked this fact. Our experimental approach (applying some
previous programs devoted to spectral sequences) allowed us
to detect a small error presented in a well-known book on
Computational Topology \cite{EH10}, and eventually make explicit
the very relation about persistence and spectral sequences.

As a consequence of this study, it was also observed that it
was not necessary to rewrite code for the persistent homology
case, neither apply the formula relating it with spectral sequences.
It is enough to modify slightly our spectral sequence programs to
compute also persistent homology.

As a by-product, since our spectral sequence algorithms were designed
for non-finite spaces (but with finitely generated homology groups;
see in Section~\ref{sec:spct-seq-infinite} the key notion of \emph{effective homology}) our
persistent homology programs can be applied also over spaces of
infinite nature (a situation which, up to our knowledge, has not
been previously solved). Similarly, our programs compute the
\emph{integer} persistent homology groups and their corresponding
\emph{integer} barcodes.

As for applications, in this paper we have computed integer persistence of the Klein bottle and the torus, and we have shown which are the integer barcodes
on a concrete Postnikov tower. Persistent homology can also be applied to biomedical
image processing (see in \cite{HPR12} a description of the kind of biological problems
where we are planning to apply persistent homology techniques); our enriched spectral sequence
programs will allow us to undertake this task.

\bibliographystyle{amsplain}
\bibliography{biblio}

Ana Romero. Departamento de Matem\'aticas y Computaci\'on, Universidad de La Rioja, Spain. E-mail: ana.romero@unirioja.es.

J\'onathan Heras. School of Computing, University of Dundee, UK. \\E-mail: jonathanheras@computing.dundee.ac.uk.

Julio Rubio. Departamento de Matem\'aticas y Computaci\'on, Universidad de La Rioja, Spain. E-mail: julio.rubio@unirioja.es.

Francis Sergeraert. Institut Fourier, Universit\'e Joseph Fourier, France. E-mail: Francis.Sergeraert@ujf-grenoble.fr.
\end{document}